\documentclass[journal=gmj]{CUP-JNL-DTM}%

\usepackage{graphicx}
\usepackage{multicol,multirow}
\usepackage{amsmath,amssymb,amsfonts}
\usepackage{mathrsfs}
\usepackage{amsthm}
\usepackage{rotating}
\usepackage{appendix}
\usepackage{ifpdf}
\usepackage[T1]{fontenc}
\usepackage{newtxtext}
\usepackage{newtxmath}
\usepackage{textcomp}
\usepackage{xcolor}
\usepackage{lipsum}
\usepackage[colorlinks,allcolors=blue]{hyperref}
\usepackage{enumitem}
\usepackage{needspace}
\usepackage{longtable}
\usepackage{booktabs}
\usepackage{array}
\usepackage{ragged2e} % better raggedright

\usepackage{tabularx}
\usepackage{booktabs}
\usepackage{array}
\usepackage{dsfont}

\theoremstyle{definition}

\newcolumntype{L}[1]{>{\RaggedRight\arraybackslash}p{#1}}
\articletype{PREPRINT MANUSCRIPT}
\jyear{2026}
\begin{document}

\begin{Frontmatter}

\title[Anchors Away: Navigating Single-Arm Indirect Comparisons with Multilevel Unanchored Meta-Regression (ML-UMR-SA)]{Anchors Away: Navigating Single-Arm Indirect Comparisons with Multilevel Unanchored Meta-Regression (ML-UMR-SA)}

\author[1]{Conor O. Chandler}
\author[2]{K. Jack Ishak}

\authormark{Chandler \textit{et al}.}

\address[1]{\orgdiv{PPD}, \orgname{Thermo Fisher Scientific}, \orgaddress{\city{Waltham}, \state{MA},  \country{United States}}. \email{conor.chandler@thermofisher.com}}

\address[2]{\orgdiv{PPD}, \orgname{Thermo Fisher Scientific}, \orgaddress{\city{Montreal}, \state{Quebec},  \country{Canada}}}

\authormark{Chandler et al.}

\keywords{population-adjusted indirect comparison; disconnected; matching-adjusted indirect comparison; health technology assessment; multilevel network meta-regression}

\abstract{Unanchored indirect treatment comparisons using single-arm evidence are increasingly used in health technology assessment (HTA) when anchored comparisons of randomized evidence are not possible. Existing population-adjusted methods, including matching-adjusted indirect comparison (MAIC) and simulated treatment comparison (STC), are generally limited to pairwise analyses and typically estimate marginal effects in the comparator study population, which may differ from the decision-relevant target population.

\hspace*{4mm} We propose multilevel unanchored meta-regression for single-arm indirect comparisons (ML-UMR-SA), a Bayesian outcome-regression framework for synthesizing individual patient data and aggregate data from single-arm evidence. ML-UMR-SA respecifies multilevel network meta-regression to pool absolute outcomes rather than relative treatment effects, jointly modeling individual- and aggregate-level data and linking aggregate outcomes through marginalization over study-specific covariate distributions, enabling estimation across multiple treatments, studies, and target populations.

\hspace*{4mm} ML-UMR-SA distinguishes assumptions required to identify treatment effects from those required to transport results to target populations. Valid inference relies on strong and often unverifiable assumptions, particularly conditional constancy of absolute effects (including no unobserved confounding and correct outcome-model specification) and cross-treatment identifying assumptions such as the shared prognostic factor assumption (SPFA). These requirements are made explicit within a Bayesian modeling framework that facilitates sensitivity analyses.

\hspace*{4mm} In simulations without unobserved confounding, ML-UMR-SA produced low bias and nominal coverage for comparator-population effects. Transportability to alternative populations depended critically on identifying assumptions: violations of SPFA led to bias under strong effect modification, whereas incorporating subgroup information restored near-unbiased estimation and nominal coverage. Results were robust to moderate covariate correlation misspecification across the simulated data-generating mechanisms considered here.}
\end{Frontmatter}

\hypertarget{introduction}{%
\section{Introduction}\label{introduction}}

Network meta-analysis (NMA) provides a principled framework for indirect
treatment comparisons (ITCs) when evidence networks are connected
through a common comparator.\textsuperscript{1} Valid inference relies
on the assumption that treatment effects are not modified by imbalances
in patient characteristics across trials.\textsuperscript{1-6}
Multilevel network meta-regression (ML-NMR) extends NMA in connected
networks by integrating individual patient data (IPD) and aggregate data
(AgD) within a hierarchical outcome regression framework, explicitly
adjusting for effect modification to mitigate bias arising from
population differences.\textsuperscript{7-9}

In many HTA applications, however, the evidence base may comprise single-arm studies and randomized controlled trials (RCTs) without a connected network of randomized comparisons, precluding an anchored comparison between treatments of interest.\textsuperscript{10-13} In certain therapeutic areas, particularly oncology, evidence for new interventions and relevant comparators are increasingly available only from single-arm trials.\textsuperscript{13-15} In such cases, decision-makers often rely on unanchored ITCs based on single-arm evidence (i.e., from single-arm trials or individual arms pulled from RCTs) without a common anchoring comparator (e.g., placebo). Despite their increasing use in HTA submissions, such analyses pose considerable challenges: observed outcome differences cannot be attributed to treatment effects rather than population or study design differences without strong, often unverifiable assumptions.\textsuperscript{16-19}

Current population-adjusted indirect comparison (PAIC) methods for
unanchored single-arm settings---most notably matching-adjusted indirect
comparison (MAIC)\textsuperscript{16,20,21} and simulated treatment
comparison (STC)\textsuperscript{16,22}---are pragmatic tools but have
important limitations. They are typically restricted to pairwise
comparisons and often estimate effects in the comparator study
population, which may not align with decision-relevant populations in
HTA, such as the index study or a jurisdiction-specific
population.\textsuperscript{19,23} More fundamentally, extending such
analyses across multiple studies and treatments, while
distinguishing assumptions needed for identification from those
additionally needed for transport to an external target population, is
not straightforward. This limitation can be understood through a causal
inference perspective: PAICs are fundamentally transportability
problems, where treatment effects estimated in one population must be
transported to a decision-relevant target
population.\textsuperscript{23} For fully unanchored single-arm
comparisons based on absolute outcomes, transport to an external
population requires the outcome model and measured covariates to be
sufficient to predict absolute outcomes in that population, imposing
assumptions beyond those required for estimation in the observed study
populations.

To address these limitations, we introduce multilevel unanchored meta-regression for single-arm indirect comparisons (ML-UMR-SA), a Bayesian outcome-regression approach for PAICs based on single-arm evidence, which may comprise collections of single-arm studies or individual treatment arms from RCTs. ML-UMR-SA jointly synthesizes IPD and AgD by modeling the relationship between outcomes, treatments, and relevant covariates, and uses this model to estimate adjusted treatment-specific outcomes and relative effects in explicitly defined target populations relevant to the decision problem. ML-UMR-SA can be viewed as a re-specification of ML-NMR for fully unanchored single-arm comparisons and as a generalization of unanchored pairwise STC.\textsuperscript{24}
In its current formulation, ML-UMR-SA produces unanchored comparisons by modeling outcomes within treatment arms as conditionally independent contributions to the likelihood, regardless of their original study design. Importantly, ML-UMR-SA does not relax the strong assumptions or overcome the inherent limitations of unanchored single-arm PAICs (Appendix A); rather, its contribution lies in extending outcome-regression-based comparisons to multiple treatment arms and studies while enabling estimation of both marginal and conditional effects in explicitly defined target populations.

This paper makes three contributions. First, we formalize the identification and transportability assumptions required for valid inference in fully unanchored comparisons based on single-arm evidence, clarifying their role in HTA practice. Second, we develop ML-UMR-SA, which integrates IPD and AgD across multiple treatment arms and studies and generalizes existing PAIC methods beyond pairwise settings. Third, through simulation, we demonstrate that in PAICs, comparator-population effects may appear robust under certain conditions, whereas valid transport to different decision-relevant populations depends critically on additional identifying assumptions, highlighting the importance of explicitly modeling and assessing sensitivity to these assumptions in practice. This work expands on our initial conference presentation of ML-UMR-SA,\textsuperscript{24}
providing the full methodological formulation, extensions to multiple treatments, studies, and survival outcomes, and simulation evaluation against existing PAIC methods. The identification and transportability framework presented herein builds on our previous estimand-based work, which examines how effect modification, collapsibility, and estimand choice influence the transportability properties of PAIC-derived treatment effects across populations.\textsuperscript{23}

\hypertarget{pairwise-formulation-of-ml-umr}{%
\section{Pairwise Formulation of
ML-UMR-SA}\label{pairwise-formulation-of-ml-umr}}

ML-UMR-SA is a Bayesian modeling approach for PAICs comprising fully disconnected single-arm evidence, which may originate from single-arm studies or individual treatment arms of randomized trials. It uses the multilevel outcome regression and marginalization framework of ML-NMR but adapts this framework specifically to unanchored comparisons based on treatment-arm-specific absolute outcomes, rather than randomized treatment contrasts. Therefore, ML-UMR-SA can be viewed as an adaptation of ML-NMR for arm-based unanchored evidence synthesis and as a principled generalization of unanchored STC. ML-UMR-SA jointly models IPD and AgD within a unified outcome regression framework, and, unlike STC, incorporates aggregate outcomes directly through a likelihood derived by relating the regression model to the marginal outcome. This shared regression structure enables marginalization over study-specific covariate distributions and transport of treatment effects to decision-relevant populations. Although ML-UMR-SA may incorporate individual treatment arms from RCTs as single-arm evidence, this would generally be considered when the available randomized comparisons are insufficient to support the indirect comparison of interest, as may also occur with unanchored MAIC and STC. In such cases, the model treats each arm as a separate contribution to the likelihood, rather than modeling within-trial randomized treatment contrasts or accounting for dependencies between arms from the same study. Therefore, anchored methods that appropriately preserve randomization should be preferred whenever feasible.

Identification relies on strong assumptions, including conditional exchangeability of potential outcomes across studies and correct specification of the outcome regression model (see Appendix A). Here, ``identification'' refers to the ability to separate treatment effects from differences in patient populations and study design across studies contributing single-arm evidence. In unanchored settings, observed differences in outcomes across studies cannot be attributed to treatment effects without additional assumptions. In unanchored population adjustment, this conditional exchangeability requirement is commonly referred to as conditional constancy of absolute effects: conditional on the included covariates, absolute outcomes under each treatment must be comparable across the relevant study populations. This requires that all important prognostic factors and effect modifiers that differ across studies are adequately accounted for and that the outcome model is correctly specified. Conditional constancy is a particularly strong and central assumption in unanchored comparisons based on single-arm evidence and cannot be directly verified from the observed data. Its plausibility should therefore be assessed using clinical and subject-matter knowledge and, where feasible, predictive checks, external validation, and sensitivity analyses. The identification assumptions are described formally in Appendix A.

To position ML-UMR-SA relative to existing PAIC methods and to ML-NMR, Table 1 summarizes key features and differences in evidence structure, modeling approach, estimand, and transportability.

\vspace{4mm}

\noindent \textbf{Table 1} Positioning ML-UMR-SA relative to STC, MAIC, and ML-NMR

\noindent \includegraphics[
    scale=.75,
    trim=0.75cm 13.5cm 0.1cm 2cm,
    clip
]{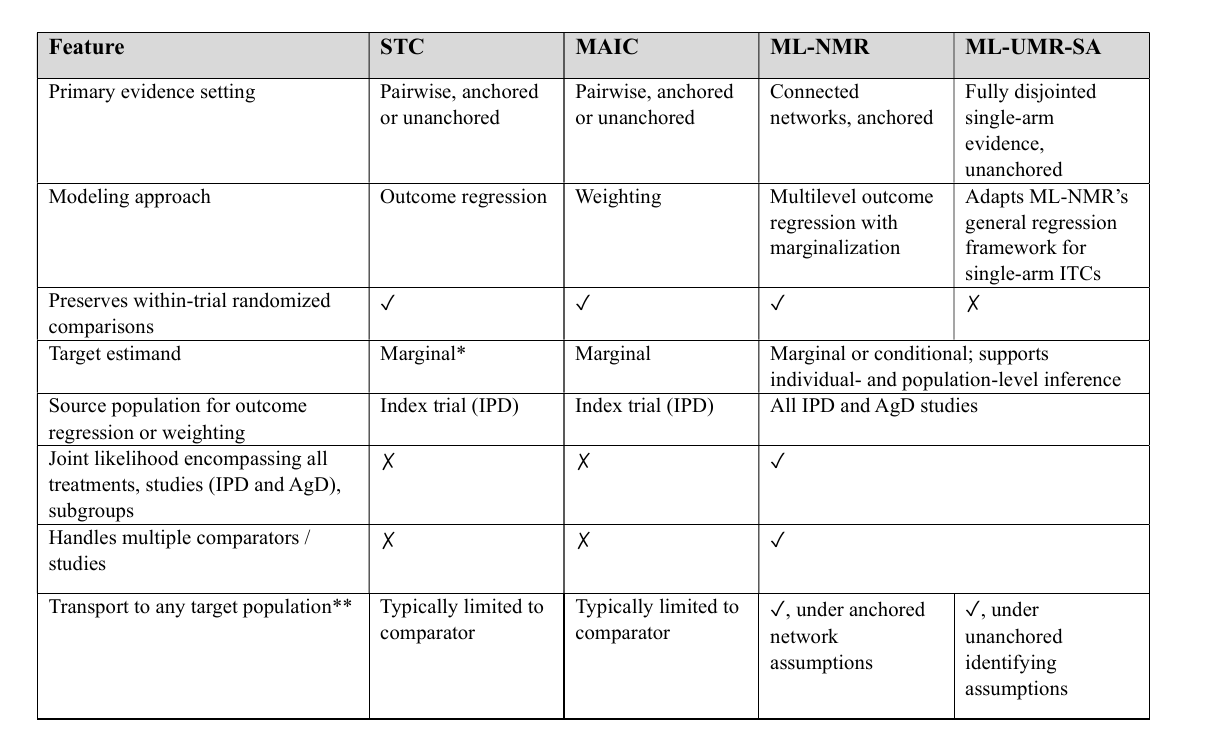}

{\footnotesize
\setlength{\baselineskip}{0.9\baselineskip}
\noindent Abbreviations: AgD = aggregate data; IPD = individual patient data; MAIC
= matching-adjusted indirect comparison; ML-NMR = multilevel network
meta-regression; ML-UMR-SA = multilevel unanchored meta-regression for
single-arm comparisons; STC = simulated treatment comparison

*For STC, marginal estimands are typically used to align with the
aggregate outcomes reported by comparator studies. Conditional estimands
may be estimated only if sufficient comparator information is available
to support a compatible conditional comparison.

**For MAIC and STC, transport is typically restricted to the comparator
study population due to data asymmetry. Under certain assumptions and
collapsible effect measures, the effect for the active-active contrast
may be invariant across populations and directly transportable to any
population. However, this is not true in general. ML-NMR and ML-UMR-SA
rely on model-based standardization, which is only valid when
identifying assumptions are met.\par}

\vspace{4mm}

To introduce ML-UMR-SA, we first consider a pairwise unanchored comparison between two single-arm trials evaluating treatment A (index), observed with IPD, and treatment B (comparator), observed with AgD only. This setup reflects the standard HTA scenario addressed by unanchored MAIC or STC. Extensions to multiple treatments and studies follow naturally within the same framework (Sections 3.1-3.2).

\hypertarget{outcome-regression-modeling}{%
\subsection{Outcome Regression
Modeling}\label{outcome-regression-modeling}}

Outcome regression underpins several PAIC approaches, including ML-UMR-SA,
ML-NMR, and STC, by specifying the relationship between treatment,
baseline covariates, and outcomes. Let \(x \in \mathbb{R}^{P}\ \)denote
a vector of baseline characteristics (prognostic factors and potential
effect modifiers). For treatment \(k \in \{ A,B\}\), consider the
generalized linear model:

\begin{equation}
g\!\left(\theta_k(x)\right)
=
\alpha_k + x^\top \beta_k
\label{eq:outcome_model}
\end{equation}

where \(g( \cdot )\) is a link function,
\(\theta_{k}(x) = \mathbb{E}\left\{ Y\  \right|\ X = x,k\}\) is the
conditional mean outcome under treatment \(k\), \(\alpha_{k}\) is a
treatment-specific intercept or baseline, and
\(\beta_{k} \in \mathbb{R}^{P}\) represent \(P\) covariate effects under
treatment \(k\). The regression coefficients may be treatment-specific
or constrained to be common across treatments, depending on the
identifying assumptions adopted. Differences between treatment-specific
coefficients (\(\beta_{A} \neq \beta_{B}\)) represent treatment effect
modification on the linear predictor scale, whereas constraining
coefficients to be common across treatments corresponds to assuming the covariates are not modifiers of the effect of $A$ vs. $B$ on the linear predictor scale—i.e., they share common prognostic effects on the outcome. However, unlike anchored ML-NMR, the type of
covariate effect cannot generally be distinguished or separately
identified from fully disconnected evidence without additional
assumptions or information.

Unlike standard ML-NMR, which parameterizes treatment effects relative
to a reference treatment and distinguishes prognostic effects from
treatment--covariate interactions, ML-UMR-SA directly models
treatment-specific absolute outcomes. In the absence of a connected
evidence network, relative treatment effects and consistency equations
do not enter the model specification. Instead, treatment comparisons are
derived from the fitted treatment-specific outcome models after
standardization to a common target population. Consequently, the
distinction between prognostic factors and effect modifiers cannot be
identified from the disconnected single-arm evidence alone, and
identifying treatment-specific covariate effects requires additional
assumptions or information (Section 2.2). In this respect, the absolute
outcome parameterization of ML-UMR-SA is conceptually related to
arm-based meta-analysis, which synthesizes absolute outcomes rather than
relative treatment effects.\textsuperscript{25} However, ML-UMR-SA
extends this approach for adjusted single-arm ITCs through
individual-level covariate adjustment and marginalization over
study-specific covariate distributions.

A further distinction from standard ML-NMR concerns prediction in
external target populations. In anchored ML-NMR, relative treatment
effects can be combined with an observed or externally informed baseline
risk to obtain absolute outcome predictions. In contrast, ML-UMR-SA
models treatment-specific absolute outcomes directly, with target
population predictions obtained by marginalizing the fitted outcome
models over the target covariate distribution. Consequently, an observed
baseline risk in an external population cannot be used to calibrate
these predictions through a reference-treatment model in the same manner
as in anchored ML-NMR. This places greater reliance on the
transportability of the absolute outcome models, including their
baseline parameters, across study populations.

At the individual level, the conditional distribution of the outcome of
interest, \(Y\), is generally expressed as:

\begin{equation}
Y \mid X, k \sim \pi\!\left(\theta_k(x)\right),
\label{eq:outcome_distribution}
\end{equation}

where \(\pi( \cdot )\) denotes an appropriate distribution (e.g.,
Bernoulli, Normal, or Poisson).As in ML-NMR, the conditional outcome
distribution and corresponding link function are selected according to
the outcome of interest, accommodating continuous, binary, count, and
rate outcomes; extensions to time-to-event outcomes using general
likelihoods are described in Section 3.3.

This regression model forms the basis of ML-UMR-SA, integrating
contributions from both IPD and AgD within a common outcome-modeling
framework. Identifying assumptions that constrain covariate effects to
be common across treatments allow information on these effects to be
shared across otherwise disconnected treatment evidence.

\hypertarget{individual-level-component-treatment-a}{%
\subsubsection{Individual-Level Component (Treatment
A)}\label{individual-level-component-treatment-a}}

For treatment \(A\), suppose individual outcomes are observed for
individuals \(i = 1,\ldots,n_{A}\), with covariates \(x_{iA}\). Let
\(Y_{iA}\) denote the observed outcome for individual \(i\) under
treatment \(A\).

The individual-level model is:

\[
g(\theta_{iA}) = \alpha_A + x_{iA}^{\top}\beta_A
\]

where
\(\theta_{iA}\mathbb{= E}\left\{ Y_{iA}\mid X_{iA} = x_{iA},k = A \right\} = g^{- 1}(\alpha_{A} + x_{iA}^{\top}\beta_{A})\)
represents the conditional mean outcome and
\(Y_{iA}\ |\ X_{iA} \sim \pi_{\text{IPD}}\left( \theta_{iA} \right)\).
The intercept \(\alpha_{A}\) captures the baseline outcome level in the
study evaluating treatment \(A\) after adjustment for covariates. The
individual-level likelihood contribution is:

\[
\mathcal{L}_{\mathrm{IPD}}(\alpha_A,\beta_A)
=
\prod_{i=1}^{n_A}
\pi_{\mathrm{IPD}}
\!\left(
y_{iA}\mid x_{iA};\alpha_A,\beta_A
\right)
\]

This component informs the estimation of the covariate--outcome
relationship for treatment A, yielding posterior inference for
\(\left( \alpha_{A},\beta_{A} \right)\). In the unanchored setting,
these IPD provide the only individual-level information on the
association between baseline covariates and outcomes.

\hypertarget{aggregate-level-component-treatment-b}{%
\subsubsection{Aggregate-Level Component (Treatment
B)}\label{aggregate-level-component-treatment-b}}

Suppose a single study evaluates treatment \(B,\) and only AgD are
available. Individual outcomes \(Y_{iB}\) and covariates \(X_{iB}\) are
not observed. For instance, covariate summaries may include means,
standard deviations, or medians, while outcomes may be reported as a
sample mean \(\bar{y}_B\) (continuous outcome) or event
counts out of \(n_{B}\) individuals (binary outcome).

The key challenge is that the regression model in (1) is specified at
the individual level, whereas the observed data for treatment B are
aggregated over the study population. To link the individual-level
outcome regression model to the observed AgD for treatment B, the
likelihood must be expressed in terms of the marginal outcome, obtained
by integrating the conditional outcome model over the covariate
distribution in the comparator population.

\hypertarget{marginalization-over-the-study-specific-covariate-distribution}{%
\paragraph{Marginalization over the Study-Specific Covariate
Distribution}\label{marginalization-over-the-study-specific-covariate-distribution}}

Under the outcome regression model, the conditional mean outcome for an
individual with covariates \(x\) under treatment \(B\) is

\[
g\!\left(\theta_B(x)\right)
=
\alpha_B + x^\top \beta_B.
\]

Let \(f_{B}(x)\) denote the joint distribution of baseline covariates in
the comparator study population. The marginal mean outcome is

\begin{equation}
\theta_{\bullet B}
=
\mathbb{E}_{X \sim f_B}\!\left\{\theta_B(X)\right\}
=
\int \theta_B(x)\,f_B(x)\,dx
\label{eq:marginal_mean}
\end{equation}

Substituting the regression model yields

\begin{equation}
\theta_{\bullet B}
=
\int g^{-1}\!\left(\alpha_B + x^\top \beta_B\right)
\, f_B(x)\, dx
\label{eq:marginalized_mean}
\end{equation}

This integral represents the expected outcome under treatment \(B\),
averaged over the study-specific covariate distribution. Through the
marginalization step in (3)-(4), the outcome regression model is
effectively constrained to replicate the marginal outcome \(\theta_{B}\)
observed in the AgD study.

Marginalization as shown in (4) is essential, particularly for nonlinear
models (e.g., logistic or log links), where

\[
\theta_{\bullet B}
\neq
g^{-1}\!\left(\alpha_B + \bar{x}_B^{\top}\beta_B\right)
\]

except under an identity link. Evaluating the inverse link at the mean
covariate vector leads to aggregation bias.

The aggregate outcome is modeled as

\[
Y_{\bullet B} \mid \theta_{\bullet B}
\sim
\pi_{\mathrm{AgD}}\!\left(\theta_{\bullet B}\right)
\]

where \(\pi_{\text{AgD}}( \cdot )\) is an appropriate aggregate-level
distribution (e.g., Binomial for binary outcomes or Normal for
continuous outcomes). The corresponding aggregate-level likelihood
contribution is

\begin{equation}
\mathcal{L}_{\mathrm{AgD}}(\alpha_B,\beta_B)
=
\pi_{\mathrm{AgD}}
\!\left(
y_{\bullet B} \mid \theta_{\bullet B}
\right)
\label{eq:agd_likelihood}
\end{equation}

When all covariates are binary, the marginal mean reduces to a finite
weighted sum over covariate strata:

\begin{equation}
\theta_{\bullet B}
=
\sum_{x \in \{0,1\}^{P}}
g^{-1}\!\left(\alpha_B + x^\top \beta_B\right)
\mathbb{P}_B(X=x)
\label{eq:discrete_marginalization}
\end{equation}

where \(\mathbb{P}_{B}(X = x)\) denotes the joint probability of
covariate pattern \(x\) in the comparator population. In (6),
marginalization corresponds to model-based standardization over the
fully enumerated discrete covariate distribution. Numerical integration
may be unnecessary when the joint distribution is known and can be
enumerated explicitly. However, when only marginal summaries are
available, there are a large number of covariates, or when covariates
are continuous, the integral in (4) must be approximated numerically.

\hypertarget{reconstruction-of-the-covariate-distribution-f_bx}{%
\paragraph{\texorpdfstring{Reconstruction of the Covariate Distribution
\(f_{B}(x)\)}{Reconstruction of the Covariate Distribution f\_\{B\}(x)}}\label{reconstruction-of-the-covariate-distribution-f_bx}}

The evaluation of the marginal mean in (4) requires specification of the
joint covariate distribution \(f_{B}(x)\), which is not directly
observed. Although marginal summaries are typically reported,
correlations between covariates are rarely available. The joint
distribution must, therefore, be approximated under additional
assumptions before fitting the aggregate-level component of ML-UMR-SA.

ML-UMR-SA adopts the same principled strategy as ML-NMR
straightforwardly: 1) specify marginal distributions consistent with
reported summaries, 2) specify or estimate the dependence structure
between covariates (e.g., via correlations informed by IPD or external
data), and 3) combine marginals with the assumed dependence structure to
construct an approximate joint distribution suitable for quasi--Monte
Carlo integration (see Appendix B).\textsuperscript{8}

This yields a set of \(Q\) integration points
\(\left\{ x_{qB}^{*} \right\}_{q = 1}^{Q}\), where each
\(x_{qB}^{*}\ \)is drawn from the approximated joint distribution,
providing a numerical representation of \(f_{B}(x)\), the unobserved
joint covariate distribution in the comparator population. This provides
a practical approximation to \(f_{B}(x)\) consistent with reported
marginal summaries and an assumed dependence structure; however,
inference may be sensitive to these assumptions, particularly when
correlations are misspecified or covariate information is limited.

\hypertarget{numerical-approximation-of-the-marginal-mean}{%
\paragraph{Numerical Approximation of the Marginal
Mean}\label{numerical-approximation-of-the-marginal-mean}}

Given integration points \(x_{qB}^{*}\), the marginal mean in (3)-(4) is
approximated as

\[
\theta_{\bullet B}
\approx
\frac{1}{Q}
\sum_{q=1}^{Q}
g^{-1}\!\left(
\alpha_B + (x_{qB}^{*})^\top \beta_B
\right),
\]

where \(Q\) denotes the number of quasi--Monte Carlo samples.
Quasi--Monte Carlo integration improves numerical efficiency relative to
standard Monte Carlo sampling.\textsuperscript{26}

In the Bayesian implementation, this integration is performed at each
posterior draw of \(\left( \alpha_{B},\beta_{B} \right)\), ensuring full
propagation of parameter uncertainty into marginal predictions and
treatment effects.

\hypertarget{joint-likelihood}{%
\subsubsection{Joint Likelihood}\label{joint-likelihood}}

The full joint likelihood for ML-UMR-SA is:

\begin{equation}
\mathcal{L}_{\mathrm{joint}}
=
\mathcal{L}_{\mathrm{IPD}}(\alpha_A,\beta_A)
\times
\mathcal{L}_{\mathrm{AgD}}(\alpha_B,\beta_B)
\label{eq:joint_likelihood}
\end{equation}

or equivalently,

\begin{equation}
\log\!\left(\mathcal{L}_{\mathrm{joint}}\right)
=
\log\!\left(
\mathcal{L}_{\mathrm{IPD}}(\alpha_A,\beta_A)
\right)
+
\log\!\left(
\mathcal{L}_{\mathrm{AgD}}(\alpha_B,\beta_B)
\right)
\label{eq:log_joint_likelihood}
\end{equation}

The two components in (7)-(8) are conditionally independent given the
model parameters. When the identifying assumptions impose shared
regression parameters (Section 2.2), information is shared between
components through these common parameters.

This joint multilevel regression structure enables coherent estimation
across treatments and supports transport of treatment effects to target
populations. However, valid causal interpretation continues to rely on
identifying assumptions, including cross-study conditional
exchangeability and correct model specification.

\hypertarget{shared-prognostic-factor-assumption}{%
\subsection{Shared Prognostic Factor
Assumption}\label{shared-prognostic-factor-assumption}}

In the disconnected single-arm setting, differences in observed outcomes
depend on treatment, population, and study design and cannot be
disentangled without structural assumptions on the outcome model. As
described in Section 2.1, the treatment-specific coefficients
\(\beta_{k}\) represent covariate effects under each treatment. In the absence of a common reference treatment, these coefficients cannot generally be decomposed into prognostic and effect-modifying components (relative to a reference arm) from the available evidence alone. For identification, we consider the shared prognostic factor assumption (SPFA), under which all included covariate effects are assumed to be shared across treatments:

\[\beta_{A} = \beta_{B} = \beta.\]

That is, all included covariates are assumed to have the same effects
across treatments \(A\) and \(B\). Consequently, these covariates act
prognostically rather than as effect modifiers for the \(A\) vs. \(B\)
contrast on the linear predictor scale, with treatment-specific
differences on this scale captured through the intercept parameters
\(\alpha_{A}\) and \(\alpha_{B}\). Thus, SPFA implies the absence of
differential treatment--covariate interactions (i.e., no effect
modification is assumed) between \(A\) and \(B\) (typically the active
investigational treatments being compared) on the linear predictor
scale. Importantly, this does not imply an absence of effect
modification relative to other treatments, such as placebo or standard
of care (denoted \(C\)). The terminology is analogous to the shared
effect modifier assumption (SEMA) used in anchored PAICs but reflects
the different identifying structure of disconnected single-arm evidence:
whereas SEMA concerns sharing treatment-covariate interaction effects
relative to a common comparator, SPFA constrains the treatment-specific
covariate effects themselves to be common. Under SEMA,
effect-modification parameters shared across active treatments \(A\) and
\(B\) relative to a common comparator \(C\) cancel in the conditional
active--active contrast, yielding a constant conditional treatment
effect for \(A\) vs. \(B\) on the linear predictor scale. Similarly,
under SPFA, the shared covariate effects cancel in the conditional \(A\)
vs. \(B\) contrast, yielding a constant conditional contrast between the
single-arm treatments being compared. However, unlike SEMA, SPFA imposes
this restriction directly on absolute-outcome covariate effects.

The plausibility of SPFA should be informed by clinical and biological
considerations, including treatment mechanisms; all else equal, shared
covariate effects may be more plausible for treatments with similar
mechanisms of action or biological targets, and less plausible for
treatments acting through substantially different pathways. Importantly,
apparent departures from SPFA cannot generally be attributed uniquely to true
treatment effect modification. Estimated differences between \(\beta_{A}\) and
\(\beta_{B}\) may also reflect residual confounding or other unmeasured
differences between studies correlated with the included covariates.
These mechanisms cannot be disentangled from the disconnected single-arm
evidence without additional information.

When SPFA is invoked, IPD from treatment \(A\) contribute directly to
the estimation of the shared covariate effects \(\beta\), thereby
enabling the identification of marginal outcomes and treatment contrasts
for treatment \(B\) despite the absence of comparator IPD. The SPFA may
be relaxed to allow treatment-specific covariate effects (i.e.,
\(\beta_{A} \neq \beta_{B}\)), thereby allowing the outcome model to
accommodate differential covariate effects across treatments, including
those arising from treatment effect modification.

\hypertarget{relaxing-spfa}{%
\subsubsection{Relaxing SPFA}\label{relaxing-spfa}}

Relaxing SPFA requires additional information beyond overall aggregate
outcomes for the comparator to identify treatment-specific covariate
effects.

One approach is to incorporate subgroup-level AgD for treatment B. For
subgroups \(s = 1,\ldots,S\) in the comparator study, the aggregate
likelihood contribution is written separately for each subgroup.

The subgroup-level aggregate outcome is modeled as

\[
Y_{\bullet Bs}
\sim
\pi_{\mathrm{AgD}}\!\left(\theta_{\bullet Bs}\right)
\]

where the marginal mean within subgroup \(s\) is

\[
\theta_{\bullet Bs}
=
\int
g^{-1}\!\left(\alpha_B + x^\top \beta_B\right)
\, f_{Bs}(x)\, dx
\]

Therefore, the aggregate-level likelihood contribution for subgroup
\(s\) is

\[
L_{\mathrm{AgD},s}
=
\pi_{\mathrm{AgD}}
\!\left(
y_{\bullet Bs}\mid\theta_{\bullet Bs}
\right)
\]

and under subgroup partitioning, the comparator likelihood in (5)
becomes

\[L_{\text{AgD}} = \prod_{s = 1}^{S}L_{\text{AgD},s}.\]

When subgroups are jointly defined (i.e., mutually exclusive and
collectively exhaustive; \(\sum_{s}^{}{n_{Bs} = n_{B}}\)), these
subgroup-specific likelihood contributions enable the identification of
treatment-specific covariate effects \(\beta_{B}\). This approach is
analogous to the subgroup likelihood formulation described by Phillippo
(2019) for ML-NMR, in which aggregate outcomes reported factorially for
each distinct combination of subgroup variables are incorporated through
separate subgroup-specific likelihood contributions.\textsuperscript{27}
Treatment-specific covariate effects may be modeled either independently
or as exchangeable under a hierarchical prior. The latter is useful when
covariate effects are believed to be similar, but not identical, across
treatments. Letting \(\beta_{k,s}\) denote the coefficient for covariate
\(s\) under treatment \(k\), exchangeability across treatments may be
modeled as

\[\beta_{k,s}\mathcal{\sim N(}\mu_{s},\sigma_{s}^{2}),\]

where \(\mu_{s}\) represents the average effect of covariate \(s\)
across treatments, and \(\sigma_{s}^{2}\) captures between-treatment
heterogeneity, with prior distributions specified for hyperparameters
\(\mu_{s}\) and \(\sigma_{s}^{2}\).

In practice, however, marginal subgroups are more commonly reported than
joint subgroups. The joint-subgroup likelihood above is not generally
directly applicable when outcomes are reported marginally by separate
subgroup variables because the resulting subgroup outcomes are
correlated. Network meta interpolation (NMI)\textsuperscript{28},
another PAIC method, treats marginal subgroups as independent in a
multivariable regression, which may underestimate uncertainty by
effectively double-counting patients across overlapping subgroups.
Similar caution is required if incorporating marginal subgroup data
within ML-UMR-SA.

When only marginal subgroup data are available, SPFA may be relaxed
univariately by allowing treatment-specific effects for individual
regression parameters while constraining others to remain shared. This
permits assessment of which covariate effects materially differ across
treatments without jointly treating overlapping marginal subgroup
outcomes as independent likelihood contributions. More generally, the
Bayesian Synthetic Likelihood (BSL)-enhanced method recently developed
by Campbell et al.\textsuperscript{29} for ML-NMR provides a potential
approach for accounting for dependence among marginal subgroup summaries
and may offer a means of relaxing SPFA using such data in ML-UMR-SA.

In the absence of subgroup data, SPFA can be relaxed through structured
sensitivity analysis. Deviations may be parameterized as
\(\beta_{B} = \beta_{A} + \delta\), where \(\delta\) represents
differential covariate effects between treatments and could be varied
over clinically meaningful ranges (e.g., via tipping point analyses) or
assigned an informative prior based on external evidence or expert
elicitation. Alternatively, \(\delta\) may be modeled as a random effect
(e.g., \(\delta \sim \mathcal{N}(0,\Sigma_{\delta}))\), inducing partial
pooling toward the SPFA while allowing for treatment-specific covariate effects (closely related to hierarchical formulations on
\(\beta_{k}\) but parameterized relative to a reference treatment rather
than a common mean). This approach introduces an additional prior-driven
structure and does not fully resolve non-identifiability in disconnected
settings; inference may therefore be sensitive to assumptions on
\(\Sigma_{\delta}\). Such sensitivity analyses make explicit the degree
of effect modification required to materially alter treatment
conclusions, while acknowledging the limits of identification in
disconnected single-arm settings.

\hypertarget{transport-effects-to-target-population}{%
\subsection{Transport Effects to Target
Population}\label{transport-effects-to-target-population}}

Once treatment effects have been identified under the modeling
assumptions described above, ML-UMR-SA enables their evaluation in any
decision-relevant target population through population-specific
standardization. Transportability is achieved by integrating the
estimated outcome model over the covariate distribution of the target
population. Specifically, treatment effects in target population
\(\widetilde{P}\) are obtained by standardizing over its covariate
distribution \(f_{\widetilde{P}}(x)\). Transportability is a distinct
step from identification, allowing effects to be evaluated across
different decision contexts. When the target population is external to
the included studies, valid transport requires the fitted conditional
absolute-outcome models to remain applicable in that population. Thus,
the measured covariates and outcome model must be sufficient to predict
absolute outcomes both across the included studies and in the
external target population. This extends the conditional constancy
requirement beyond the included evidence base and may be particularly
difficult to justify when the target population differs substantially
from the study populations.

Because ML-UMR-SA specifies a hierarchical outcome regression model
across treatments, it supports estimation of treatment-specific outcomes
and effects at both the individual and population levels, including
conditional and marginal estimands, building on the corresponding
estimand framework developed for ML-NMR.\textsuperscript{8,30} In
contrast, approaches such as STC typically estimate marginal effects in
the comparator population via g-computation and do not naturally support
transport of outcomes for all treatments to alternative target
populations within a unified framework.\textsuperscript{31}

Transported (population-average) potential outcomes for treatment
\(k \in \{ A,B\}\) are obtained via Bayesian
g-computation\textsuperscript{8,23,30,31}:

\begin{equation}
\begin{aligned}
\mu_{k,\widetilde{P}}
&=
\mathbb{E}_{X \sim \widetilde{P}}
\!\left\{
\theta_k(X)
\right\} \\
&=
\int
g^{-1}\!\left(\alpha_k + x^\top \beta_k\right)
\, f_{\widetilde{P}}(x)\, dx
\end{aligned}
\label{eq:transported_outcome}
\end{equation}

In practice, the integral in (9) is evaluated numerically by averaging
model-based predictions over observed or simulated covariate profiles
from the target population at each posterior draw of
\(\left( \alpha_{k},\beta_{k} \right)\).

When effects are defined on the linear predictor scale,
\(\eta_{k}(x) = \alpha_{k} + x^{\top}\beta_{k}\), the population-average
conditional effect for A vs. B is the average of individual-level
conditional effects\textsuperscript{23}:

\[
\begin{aligned}
\Delta_{A,B}^{\mathrm{Cond}}(\widetilde{P})
&=
\mathbb{E}_{X \sim \widetilde{P}}
\!\left\{
\eta_A(X)-\eta_B(X)
\right\} \\
&=
\alpha_A-\alpha_B
+
\mu_{X,\widetilde{P}}^{\top}
(\beta_A-\beta_B)
\end{aligned}
\]

Under the SPFA, \(\beta_{A} = \beta_{B} = \beta\), such that

\[
\begin{aligned}
\eta_A(X)-\eta_B(X)
&=
(\alpha_A + X^\top\beta)
-
(\alpha_B + X^\top\beta) \\
&=
\alpha_A-\alpha_B
\end{aligned}
\]

which is constant in \(X\). Thus,

\[
\Delta_{A,B}^{\mathrm{Cond}}(\widetilde{P})
=
\alpha_A-\alpha_B,
\qquad \text{under SPFA}
\]

for all target populations \(\widetilde{P}\). Hence, when SPFA holds and
the estimand is defined on the linear predictor scale, the
population-average conditional effect is directly transportable and
invariant to the covariate distribution.\textsuperscript{23}

Marginal effects generally require explicit standardization because the
expectation operates on the outcome scale (i.e., contrast of
population-level outcomes from (9)):

\begin{equation}
\begin{aligned}
\Delta_{A,B}^{\mathrm{Marg}}(\widetilde{P})
&=
h\!\left(\mu_{A,\widetilde{P}}\right)
-
h\!\left(\mu_{B,\widetilde{P}}\right) \\
&=
h\!\left(
\int
g^{-1}\!\left(\alpha_A + x^\top\beta_A\right)
\,f_{\widetilde{P}}(x)\,dx
\right) \\
&\quad -
h\!\left(
\int
g^{-1}\!\left(\alpha_B + x^\top\beta_B\right)
\,f_{\widetilde{P}}(x)\,dx
\right)
\end{aligned}
\label{eq:marginal_effect}
\end{equation}

where \(h( \cdot )\) represents a transformation function to construct
the effect measure of interest (e.g., log for risk
ratios).\textsuperscript{23,32-34}

For nonlinear link functions (e.g., logit or log links), the marginal
contrasts in (10) depend on the full covariate distribution in
\(\widetilde{P}\), and must, therefore, be recomputed for each target
population of interest.\textsuperscript{23,32-34} ML-UMR-SA accommodates
this naturally by integrating over the specified
\(f_{\widetilde{P}}(x)\), allowing the estimation of standardized
outcomes and both marginal and population-average conditional treatment
effects in any decision-relevant population, under the previously
discussed identifying assumptions.

\hypertarget{applied-examples-and-implementation}{%
\subsection{Applied Examples and
Implementation}\label{applied-examples-and-implementation}}

Practical implementation and worked examples of ML-UMR-SA are provided
in the accompanying \emph{mlumr} R package.\textsuperscript{35} Package
vignettes demonstrate complete applications for binary, continuous,
count, rate, and survival outcomes, including data preparation,
reconstruction of AgD covariate distributions, model fitting and
diagnostics, subgroup-based identification when relaxing SPFA, and
standardization to target populations.\textsuperscript{36} These
examples illustrate how the available aggregate information and
identifying assumptions determine the ML-UMR-SA specification that can
be supported in practice.

Computational requirements depend on the number and size of studies and
treatments, the outcome likelihood, and computing environment. In
typical applications, models for standard binary, continuous, count, and
rate outcomes are usually fitted on the order of minutes, whereas
survival models can require substantially longer computation because
aggregate likelihood contributions require numerical approximation.

In practice, analysts should assess covariate overlap and the extent of
extrapolation by comparing the covariate distributions and support of
the IPD with the reconstructed AgD and target population distributions.
Model fit and posterior predictive checks, including comparisons of
model-implied and observed outcomes where possible, should also be
assessed. Standard MCMC convergence diagnostics should also be examined,
with further guidance provided in the accompanying fitting and
diagnostics vignette.\textsuperscript{36}

Importantly, satisfactory MCMC convergence does not imply that model
parameters are well identified by the data, particularly when SPFA is
relaxed using subgroup information. Relaxing SPFA for \(K\) covariates
requires estimation of \(K + 1\) treatment-specific parameters (an
intercept and \(K\) covariate effects) from the AgD treatment.
Identification therefore requires at least \(K + 1\) independent
subgroup outcome summaries with sufficient variation in the relevant
covariates. Consequently, relaxing SPFA becomes increasingly demanding
as the number of treatment-specific covariate effects increases,
requiring increasingly rich and informative subgroup data. Where
subgroup information is limited, analysts may therefore need to restrict
relaxation of SPFA to selected covariates or use structured sensitivity
analyses, while examining posterior uncertainty and prior sensitivity.
Practical guidance and identification diagnostics are provided in the
subgroup-identification vignette.\textsuperscript{36}

\hypertarget{extensions-of-ml-umr}{%
\section{Extensions of ML-UMR-SA}\label{extensions-of-ml-umr}}

\hypertarget{beyond-pairwise-comparisons}{%
\subsection{Beyond Pairwise
Comparisons}\label{beyond-pairwise-comparisons}}

The pairwise formulation above captures the core structure of ML-UMR-SA: a
shared regression model combining an individual-level likelihood with an
aggregate-level likelihood obtained through marginalization over
covariate distributions. This structure extends directly to settings
involving three or more treatments.

Let \(k \in \{ 1,\ldots,K\}\) index treatments. For each treatment
\(k\), data may be available at the individual level, the aggregate
level, or both (a given treatment may be informed by multiple studies,
some providing IPD and others AgD).

Define \(\mathcal{T}_{\text{IPD}} \subseteq \{ 1,\ldots,K\}\) as
treatments with IPD and
\(\mathcal{T}_{\text{AgD}} \subseteq \{ 1,\ldots,K\}\) as treatments
with AgD.

The outcome model remains:

\[
g\!\left(\theta_k(x)\right)
=
\alpha_k + x^\top \beta_k
\]

with conditional distribution \(Y|X,k \sim \pi(\theta_{k}(x))\), as in
Equations (1) and (2).

For treatments with IPD, the likelihood contribution is:

\begin{equation}
\mathcal{L}_{\mathrm{IPD}}
=
\prod_{k \in \mathcal{T}_{\mathrm{IPD}}}
\prod_{i=1}^{n_k}
\pi_{\mathrm{IPD}}
\!\left(
y_{ik}\mid\theta_{ik}
\right)
\label{eq:ipd_likelihood}
\end{equation}

where
\(\theta_{ik} = g^{- 1}\left( \alpha_{k} + x_{ik}^{\top}\beta_{k} \right).\)

For treatments with AgD, let

\[
\theta_k
=
\int
g^{-1}\!\left(\alpha_k + x^\top \beta_k\right)
\, f_k(x)\, dx
\]

denote the marginal mean under treatment \(k\), where \(f_{k}(x)\) is
the reconstructed covariate distribution for that treatment. The
aggregate likelihood contribution is

\begin{equation}
\mathcal{L}_{\mathrm{AgD}}
=
\prod_{k \in \mathcal{T}_{\mathrm{AgD}}}
\pi_{\mathrm{AgD}}
\!\left(
y_k \mid \theta_k
\right)
\label{eq:agd_likelihood_multi}
\end{equation}

The full ML-UMR-SA likelihood is the product of (11) and (12), assuming
conditional independence:

\[
\mathcal{L}_{\mathrm{joint}}
=
\mathcal{L}_{\mathrm{IPD}}
\times
\mathcal{L}_{\mathrm{AgD}}
\]

The pairwise comparison arises as a special case when \(K = 2\) and data
are asymmetrically available across treatments.

\hypertarget{synthesizing-multiple-studies-per-treatment}{%
\subsection{Synthesizing Multiple Studies per
Treatment}\label{synthesizing-multiple-studies-per-treatment}}

In many HTA settings, multiple studies may evaluate the same
treatment. As discussed in Section 2.2, treatment effects cannot be
separated from study-level differences in the synthesis of single-arm
evidence since each study evaluates only one treatment. Additional
modeling structure is, therefore, required to account for between-study
heterogeneity.

When multiple studies inform a treatment, ML-UMR-SA incorporates them
through pooling strategies that determine how baseline outcome
parameters (i.e., intercepts) are shared across studies. In the primary
specification, each study is treated independently (i.e., no pooling)
and assigned its own intercept (fixed study effects), allowing baseline
outcome parameters to vary across studies rather than constraining them
to a common treatment-level intercept. Let \(j = 1,\ldots,J\) index
studies and \(k(j) \in \{ 1,\ldots,K\}\) denote the treatment evaluated
in study \(j\), with study-specific intercept \(\alpha_{j}\ \)capturing
the study-specific baseline outcome level conditional on the modeled
covariates. The linear predictor becomes

\[
\eta_{ij}
=
\alpha_j + x_{ij}^{\top}\beta_{k(j)}
\]

Treatment comparisons are then constructed by standardizing
study-specific predictions to a common target population and combining
them using prespecified weights (e.g., based on study size or precision;
see Appendix B). This may be viewed as a two-stage approach, in which
study-specific models are first estimated and the resulting standardized
predictions subsequently pooled, rather than pooling information across
studies within a single outcome model. Alternatively, analysts may
retain study-specific predictions and report separate treatment
comparisons for relevant pairs of studies or study populations rather
than pooling predictions across studies. The no-pooling specification
allows study-specific intercepts to differ. However, this does not
remove the assumptions required to compare absolute outcomes across
studies: each indirect comparison still requires conditional constancy
of absolute effects across the study populations being compared.

Alternative pooling strategies include complete pooling (treatment-level
intercepts) and hierarchical models. While these approaches can improve
precision, they introduce different restrictions on residual
between-study heterogeneity in baseline parameters. Complete pooling
assumes no residual heterogeneity between studies evaluating the same
treatment ---a strong assumption that may be difficult to justify in
practice. Hierarchical models with study-level random intercepts (i.e.,
partial pooling) offer a compromise by allowing residual heterogeneity
in baseline parameters around a treatment-level mean, but, in sparse
settings, may be weakly identified and sensitive to prior specification,
particularly when few studies inform each treatment. All three pooling
strategies nevertheless retain the strong conditional constancy
assumptions required to compare absolute outcomes across disconnected
single-arm evidence and, when transporting estimates to an external
population, the corresponding transportability requirements; they differ
in how study-specific baseline parameters and residual heterogeneity are
modeled. Full formulations are provided in Appendix B.

\hypertarget{comparisons-of-survival-outcomes}{%
\subsection{Comparisons of Survival
Outcomes}\label{comparisons-of-survival-outcomes}}

Adapting ML-UMR-SA to survival outcomes involves extending the ML-NMR
framework for general likelihoods developed by Phillippo et
al.\textsuperscript{9} to the unanchored, arm-based setting. As in
Section 2, the outcome model is defined at the individual level, with
baseline covariates \(X\) entering through the linear predictor of a
parametric survival model. Let \(j\) index studies and \(k(j)\) denote
the treatment evaluated in study \(j\). For instance, a proportional
hazards model for covariate effects within each study may be written as

\[\log(h_{j}(t_{i} \mid x_{i})) = \log(h_{0j}(t_{i})) + x_{ij}^{\top}\beta_{k(j)},\]

where \(h_{0j}(t)\) is the baseline hazard for study \(j\) evaluating
treatment \(k(j)\). Under an accelerated failure time model,

\[\log(T_{ij}) = \alpha_{j} + x_{ij}^{\top}\beta_{k(j)} + \sigma_{j}\varepsilon_{ij},\]

where \(\varepsilon_{ij}\) follows a specified distribution and
\(\sigma_{j}\) is the corresponding scale parameter. Depending on the
chosen survival distribution and available evidence, baseline, scale, or
shape parameters may be shared, partially pooled, or allowed to vary
across studies and treatments; such restrictions are modeling choices
rather than inherent requirements of ML-UMR-SA. For example, under a
Weibull specification, proportional hazards across treatments may be
imposed by sharing the shape parameter, while allowing baseline hazard
levels to differ.

For survival outcomes, the aggregate-level likelihood is generally not
available in closed form. Unlike binary or continuous outcomes, where
aggregate data can often be linked directly to a marginal mean outcome,
survival likelihoods depend on the full event-time distribution. When
only AgD are reported, individual event times and censoring indicators
are reconstructed from published Kaplan--Meier (KM) curves to
approximate the aggregate likelihood. Parametric survival models (e.g.,
Weibull) are then jointly fitted to observed IPD and reconstructed
event-time data, with marginalization over the study-specific covariate
distribution as described below.

This survival extension for ML-UMR-SA relies on additional modeling and
reconstruction steps relative to the non-survival setting, and inference
may be sensitive to additional assumptions regarding the survival model,
covariate distribution, and KM reconstruction.

\vspace{3mm}
\textbf{Conditional Likelihood (IPD)}

For study \(j\) evaluating treatment \(k(j)\) with IPD, each individual
contributes the standard conditional survival likelihood,

\[\mathcal{L}_{ij|x}^{\text{IPD}} = h_{j}(t_{i} \mid x_{i})^{\delta_{i}}S_{j}(t_{i} \mid x_{i}),\]

where \(h_{j}(t \mid x)\) and \(S_{j}(t \mid x)\) denote the conditional
hazard and survival functions for study \(j\) evaluating treatment
\(k(j)\), \(\delta_{i} \in \{ 0,1\}\) denotes the event indicator, and
\(t_{i}\) denotes the observed follow-up time.

The full IPD likelihood for study \(j\) is

\[\mathcal{L}_{j|x}^{\text{IPD}} = \prod_{i = 1}^{n_{j}}h_{j}(t_{i} \mid x_{i})^{\delta_{i}}S_{j}\left( t_{i}\mid x_{i} \right),\]

which corresponds to standard parametric survival regression. As in the
non-survival setting (Section 2), IPD identify the covariate--survival
relationship through the outcome regression model.

\vspace{3mm}
\textbf{Marginal Likelihood (AgD)}

For treatments with only aggregate survival data, individual outcome
data (event times) and covariates are unobserved. Published KM curves
can be digitized and pseudo-IPD reconstructed using methods such as the
Guyot algorithm.\textsuperscript{37} This yields estimated event times
\(t_{i}\) and event indicators \(\delta_{i}\) consistent with the
reported KM curve but does not recover individual-level covariates.

Because the survival model is specified conditionally on \(x\) and
reconstructed pseudo-IPD do not include covariates, each likelihood
contribution must be marginalized over \(f_{j}(x)\) to link the
conditional survival model to the observed aggregate survival data.

The marginal likelihood contribution for reconstructed individual \(i\)
in study \(j\) evaluating treatment \(k(j)\) is

\[
\mathcal{L}_{ij}^{\text{AgD}}
= \int h_j(t_i \mid x)^{\delta_i}
S_j(t_i \mid x)\, f_j(x)\, dx
\]

and the aggregate likelihood is:

\begin{equation}
\begin{aligned}
\mathcal{L}_{j}^{\mathrm{AgD}}
&=
\prod_{i=1}^{n_j}
\mathcal{L}_{ij}^{\mathrm{AgD}} \\
&=
\prod_{i=1}^{n_j}
\int
h_j(t_i \mid x)^{\delta_i}
S_j(t_i \mid x)
\, f_j(x)\, dx
\end{aligned}
\label{eq:agd_survival_likelihood}
\end{equation}

As no closed-form solution to (13) is typically available, this integral
is approximated numerically (e.g., quasi--Monte Carlo with copula-based
reconstruction of \(f_{j}(x)\)) at each posterior draw.

\vspace{3mm}
\textbf{Transport Survival Effects}

Following model estimation, outcomes and treatment effects can be
transported to target population \(\widetilde{P}\) (Section 2.3).
ML-UMR-SA supports estimation of both conditional and marginal survival
estimands. As shown in Section 2.3, population-average conditional
effects are typically defined directly through the parameters of the
outcome regression model. Marginal effects are obtained by standardizing
predicted survival outcomes over the covariate distribution of a target
population and are therefore population-specific.

When multiple studies inform the same treatment, study-specific survival
predictions may be standardized and combined using the pooling
strategies described in Section 3.2. Let \(S_{k}(t \mid x)\) and
\(h_{k}(t \mid x)\) denote the resulting treatment-specific conditional
survival and hazard functions used for prediction.

Population-standardized survival and hazard functions are

\begin{equation}
\begin{split}
\overline{S}_{k}\!\left(t \mid \widetilde{P}\right)
&= \mathbb{E}_{X\sim\widetilde{P}}
   \left\{S_{k}\!\left(t \mid x\right)\right\} \\
&= \int S_{k}(t \mid x)\,
   f_{\widetilde{P}}(x)\,dx .
\end{split}
\label{eq:avg-survival}
\end{equation}

and

\begin{align}
\begin{split}
\overline{h}_{k}\!\left(t \mid \widetilde{P}\right)
&=
\frac{
\mathbb{E}_{X\sim\widetilde{P}}
\left\{
h_{k}(t \mid x)\,S_{k}(t \mid x)
\right\}
}{
\mathbb{E}_{X\sim\widetilde{P}}
\left\{
S_{k}(t \mid x)
\right\}
} \\
&=
\frac{
\int h_{k}(t \mid x)\,S_{k}(t \mid x)\,
f_{\widetilde{P}}(x)\,dx
}{
\int S_{k}(t \mid x)\,
f_{\widetilde{P}}(x)\,dx
}.
\end{split}
\label{eq:avg-hazard}
\end{align}

respectively.

Marginal treatment effects are then calculated from these standardized
functions in (14) and (15). For example, the marginal hazard ratio for
\(A\) vs. \(B\) is

\begin{equation*}
\mathrm{HR}_{A,B}^{\mathrm{Marg}}\!\left(t \mid \widetilde{P}\right)
=
\frac{
\overline{h}_{A}\!\left(t \mid \widetilde{P}\right)
}{
\overline{h}_{B}\!\left(t \mid \widetilde{P}\right)
}.
\end{equation*}

Importantly, hazard ratios are non-collapsible. Consequently, even when
SPFA holds and the specified survival model additionally imposes
proportional hazards between treatments \(A\) and \(B\), such that

\begin{equation*}
\mathrm{HR}_{A,B}^{\mathrm{Cond}}
=
\frac{h_{A}(t \mid x)}{h_{B}(t \mid x)}
=
\exp\!\left(\delta_{AB}\right).
\end{equation*}

is constant over \(t\) and \(x\), the corresponding marginal hazard
ratio is generally time-varying and population-specific. In contrast,
population-average conditional treatment effects remain defined through
the survival-model parameters and, under appropriate assumptions, are
invariant to the covariate distribution of the target population,
analogous to the properties discussed in Section 2.3.

This has practical implications for economic modeling, as a single
constant marginal HR may not adequately represent treatment effects over
time even when the conditional survival model satisfies proportional
hazards. Where appropriate, the standardized treatment-specific survival
functions, \({\overline{S}}_{k}\left( t\mid\widetilde{P} \right)\), can
instead be incorporated directly into partitioned survival or other
CEMs, avoiding the need to represent treatment effects through a
constant marginal HR.

Transport of absolute survival outcomes to an external target population
requires the transportability assumption described in Section 2.3 to
extend to the full conditional survival model. Specifically, the
conditional survival model used for prediction---including relevant
baseline hazard and shape---must remain applicable in the external
population. Standardization over \(f_{\widetilde{P}}(x)\) does not
incorporate an externally observed baseline hazard; therefore,
differences in baseline survival between the included studies and target
population that are not explained by the modeled covariates may bias
transported survival estimates. As with the transport of other absolute
outcomes described in Section 2.3, this requirement may be particularly
difficult to justify when the external target population differs
substantially from the populations represented in the included studies.

A preliminary proof-of-concept simulation study evaluating ML-UMR-SA for
survival outcomes has demonstrated low bias and near-nominal coverage
for marginal survival estimands when SPFA holds, with increasing bias
and reduced coverage under violations of this
assumption.\textsuperscript{38} However, further evaluation across more
complex survival models, covariate structures, and evidence
configurations is warranted.

\hypertarget{simulation-study}{%
\section{Simulation Study}\label{simulation-study}}

We conducted a Monte Carlo simulation study to evaluate the performance
of ML-UMR-SA for unanchored indirect comparisons using disconnected
single-arm evidence. The study compared ML-UMR-SA under SPFA and with
SPFA relaxed (via subgroup data) with STC implemented via g-computation
and MAIC. For MAIC and STC, effects estimated in the comparator
population are directly transported to the index, as commonly done in
practice.\textsuperscript{23} Full simulation details and results are
provided in Appendix C, including naïve comparisons as a benchmark.
Complete numerical simulation results, including bias, empirical
standard error (SE), root mean square error (RMSE), coverage, and
interval width for each scenario and method, are provided in a CSV in
the Supplementary Material. All simulations were conducted using the R
package \emph{mlumr}.\textsuperscript{35} The complete code used to
generate and analyze the simulation study is publicly available on
GitHub.\textsuperscript{39} Models were fitted using the default prior
specifications and Stan-based MCMC implementation in \emph{mlumr}, with
three chains and 2,000 iterations per chain; additional implementation
details and guidance on model diagnostics are provided in the package
documentation and vignettes.

The data-generating mechanism comprised two independent single-arm
studies (index with IPD and comparator with AgD only) and a binary
outcome generated from a logistic regression model. Four factors were
varied in a factorial design: (i) strength of effect modification (none,
weak, or strong), (ii) degree of population imbalance between studies
(moderate vs high), (iii) covariate dependence structure (common vs
misspecified), and (iv) sample size (\(n_{A} = n_{B} = 150\) vs.
\(300\)). Each scenario used 500 Monte Carlo replications. Performance
was evaluated for marginal treatment effects in both the comparator and
index populations, focusing on log odds ratios and log risk ratios.

Across all scenarios, the naïve comparisons were substantially biased
(absolute values ranged from 0.23 to 1.63, with average bias of 0.80)
under population imbalance and exhibited poor coverage of 95\% credible
intervals, frequently \(\leq 20\%\) (Appendix C). In contrast, ML-UMR-SA,
MAIC, and STC consistently recovered effects in the comparator's
population with minimal bias and near-nominal coverage, even when SPFA
was violated and effect modification was strong (Figs. 1-4), with bias
close to zero (typically \textless0.02 on the log scale) and coverage
probabilities near 95\%.

Differences emerged when transporting effects to the index population.
For marginal log risk ratios, MAIC and STC exhibited substantial bias
(-0.16 to -0.19) and coverage of 80\%-90\% under population imbalance
even in the absence of effect modification, reflecting that such
marginal effects are not generally invariant across populations, despite
this assumption often being made in practice (Figs. 1 and 3). A smaller
degree of bias (\textless0.08) and coverage close to 95\% were observed
for log odds ratios under no effect modification (Figs. 2 and 4). In
these simulations, this lower bias reflected the greater similarity of
the true marginal log odds ratios across the index and comparator
populations relative to the corresponding log risk ratios; however, this
pattern is specific to the data-generating model and scenarios
considered and may differ under alternative outcome models or effect
scales. Due to non-collapsibility, marginal log odds ratios are not
generally invariant across populations---even in the absence of effect
modification---and transport bias may arise when effect modification or
population imbalance is present. For both log risk ratios and log odds
ratios, bias increased with the strength of effect modification and
extent of population imbalance. In the most extreme scenarios with
strong effect modification and high imbalance, bias in the index
population was substantial, reaching approximately 0.50 for log odds
ratios under MAIC or STC and approximately 0.20 for log risk ratios.
Coverage fell to around 66\%--76\%, depending on the estimand and
method; in contrast, bias in the comparator population remained close to
zero (Figs. 1-4).

ML-UMR-SA under SPFA performed well when SPFA held (i.e., no effect
modification), effect modification was weak, or population imbalance was
moderate (bias \textless0.06 and coverage 93.8\% to 97.4\%) but showed
increasing bias as violations of SPFA became more severe and population
imbalance was high, with bias up to 0.41 (Figs. 1-4). In contrast, when
SPFA was relaxed using additional identifying information (subgroup
data), ML-UMR-SA restored near-unbiased estimation (typically
\textless0.02) and nominal coverage (92\% to 96\%) across scenarios.
These patterns indicate that, in comparisons of binary outcomes,
violations of SPFA primarily affect transportability to the index
population rather than the comparator population. This parallels
simulation findings for ML-NMR with binary outcomes under violations of
the analogous SEMA, where estimates remained approximately unbiased in
the comparator population but became biased when a misspecified SEMA was
invoked to transport effects to other populations.\textsuperscript{7}

All methods were relatively robust to moderate misspecification of the
covariate dependence structure (true correlation 0.25 vs assumed 0.50),
with only minor impacts on bias and coverage. In practice, however,
performance may be more sensitive to larger discrepancies in covariate
correlations or more complex multivariate dependence structures than
those considered here.

Increasing the sample size from \(n = 150\) to \(n = 300\) did not
materially change bias (Fig. C.5), confirming that bias arises from
violations of the identifying and transportability assumptions rather
than sampling variability. However, reduced variance at larger sample
sizes led to tighter intervals and more pronounced under-coverage in
scenarios with strong effect modification and high imbalance, as
increased precision concentrates inference around biased estimates (Fig.
5). This highlights that larger sample sizes do not mitigate structural
bias and may instead make its consequences more apparent.

To assess violation of conditional constancy of absolute effects due to
an unobserved prognostic factor, we additionally repeated the
\(n = 150\) simulations while omitting \(X_{1}\) from the PAIC models
for all methods (Appendix C, Figs. C.6--C.9). Omission of this
prognostic factor resulted in bias across all methods, particularly
under greater population imbalance, with corresponding reductions in
coverage. These findings demonstrate that all methods are susceptible to
residual bias arising from unmeasured covariate differences between
study populations, emphasizing the fundamental importance of the
conditional constancy assumption in unanchored comparisons.

Overall, these simulations demonstrate that, for the binary outcomes
considered, all PAIC methods can reliably estimate effects in the
comparator population when relevant covariate differences are adequately
accounted for, even when SPFA is violated. In contrast, estimation in
the index population is more sensitive to assumptions governing
treatment-specific covariate effects: violations of SPFA led to
increasing bias with greater effect modification and population
imbalance, whereas relaxing SPFA using subgroup information
substantially reduced this bias and restored near-nominal coverage.
Finally, omission of an imbalanced prognostic factor introduced bias
across all methods and target populations, particularly under greater
population imbalance, highlighting the fundamental importance of
adequately accounting for covariate differences between study
populations.

\vspace{3mm}
\noindent \textbf{Fig. 1} Bias for Marginal Log Risk Ratio (n=150)

\includegraphics[scale=0.50]{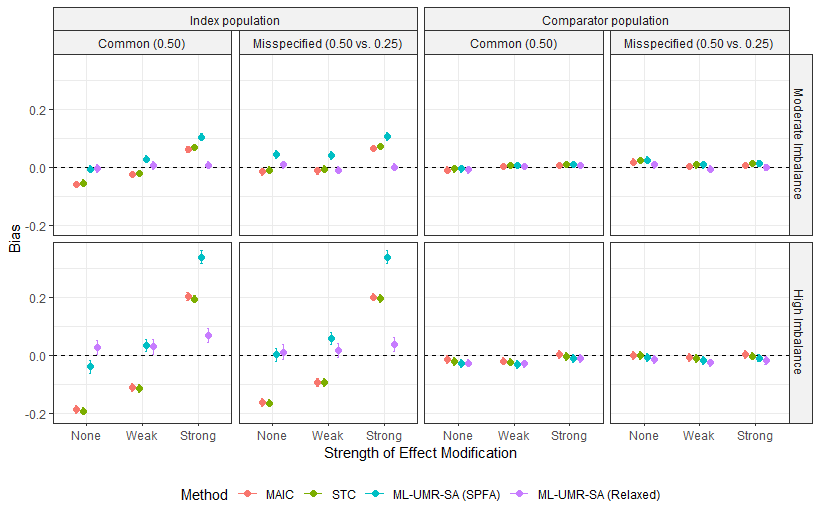}

{\footnotesize
\setlength{\baselineskip}{0.9\baselineskip}
\noindent Abbreviations: MAIC = matching-adjusted indirect comparison; ML-NMR = multilevel network meta-regression; ML-UMR-SA = multilevel unanchored meta-regression for single-arm indirect comparisons; SPFA = shared prognostic factor assumption; STC = simulated treatment comparison

Bias is shown by method and target population for scenarios varying effect-modification strength, population imbalance, and assumed comparator covariate correlation. Error bars represent ±1.96 × MCSE for each empirical bias estimate.
 \par
}

\vspace{3mm}
\noindent \textbf{Fig. 2} Bias for Marginal Log Odds Ratio (n=150)

\includegraphics[scale=0.50]{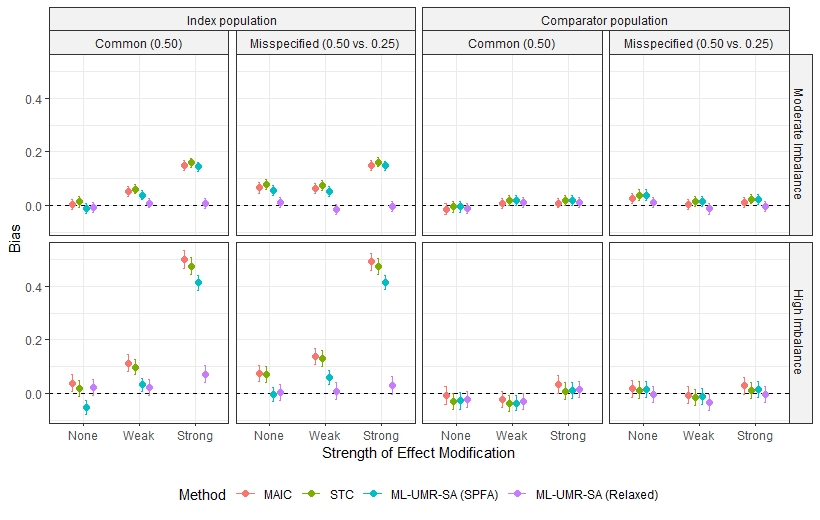}

{\footnotesize
\setlength{\baselineskip}{0.9\baselineskip}
\noindent Abbreviations: MAIC = matching-adjusted indirect comparison; ML-NMR = multilevel network meta-regression; ML-UMR-SA = multilevel unanchored meta-regression for single-arm indirect comparisons; SPFA = shared prognostic factor assumption; STC = simulated treatment comparison

Bias is shown by method and target population for scenarios varying effect-modification strength, population imbalance, and assumed comparator covariate correlation. Error bars represent ±1.96 × MCSE for each empirical bias estimate.
 \par
}

\vspace{3mm}
\noindent \textbf{Fig. 3} Coverage of 95\% Credible Intervals for Marginal Log
Risk Ratio (n=150)

\includegraphics[scale=0.50]{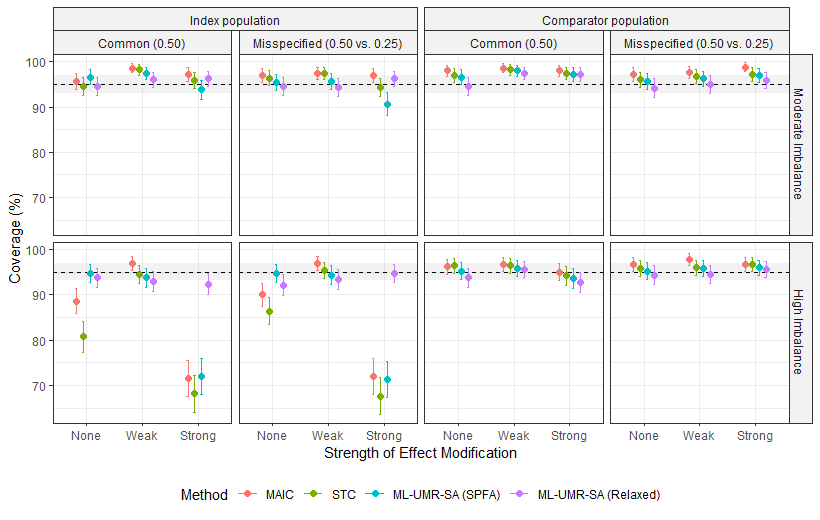}

{\footnotesize
\setlength{\baselineskip}{0.9\baselineskip}
\noindent Abbreviations: MAIC = matching-adjusted indirect comparison; MCSE =
Monte Carlo standard error; ML-NMR = multilevel network meta-regression;
ML-UMR-SA = multilevel unanchored meta-regression for single-arm
indirect comparisons; SPFA = shared prognostic factor assumption; STC =
simulated treatment comparison

Empirical coverage is shown by method and target population for
scenarios varying effect-modification strength, population imbalance,
and assumed comparator covariate correlation. Error bars represent ±1.96
× MCSE for each coverage estimate; shading indicates the Monte Carlo
reference band around nominal 95\% coverage based on 500 replicates. \par
}

\vspace{3mm}
\noindent \textbf{Fig. 4} Coverage of 95\% Credible Intervals for Marginal Log
Odds Ratio (n=150)

\includegraphics[scale=0.50]{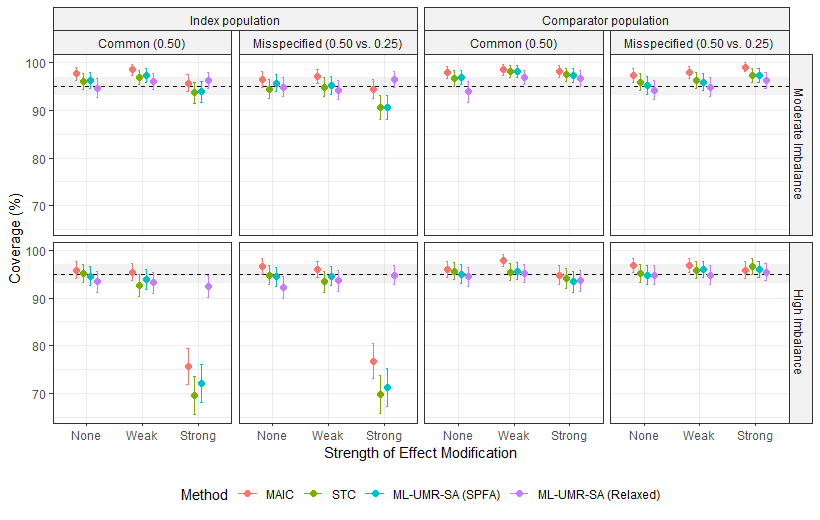}

{\footnotesize
\setlength{\baselineskip}{0.9\baselineskip}
\noindent Abbreviations: MAIC = matching-adjusted indirect comparison; MCSE =
Monte Carlo standard error; ML-NMR = multilevel network meta-regression;
ML-UMR-SA = multilevel unanchored meta-regression for single-arm
indirect comparisons; SPFA = shared prognostic factor assumption; STC =
simulated treatment comparison

Empirical coverage is shown by method and target population for
scenarios varying effect-modification strength, population imbalance,
and assumed comparator covariate correlation. Error bars represent ±1.96
× MCSE for each coverage estimate; shading indicates the Monte Carlo
reference band around nominal 95\% coverage based on 500 replicates. \par
}

\vspace{3mm}
\noindent \textbf{Fig. 5} Effect of Sample Size on Coverage Probabilities in Index
Population (Common Correlation of 0.5)

\includegraphics[scale=0.50]{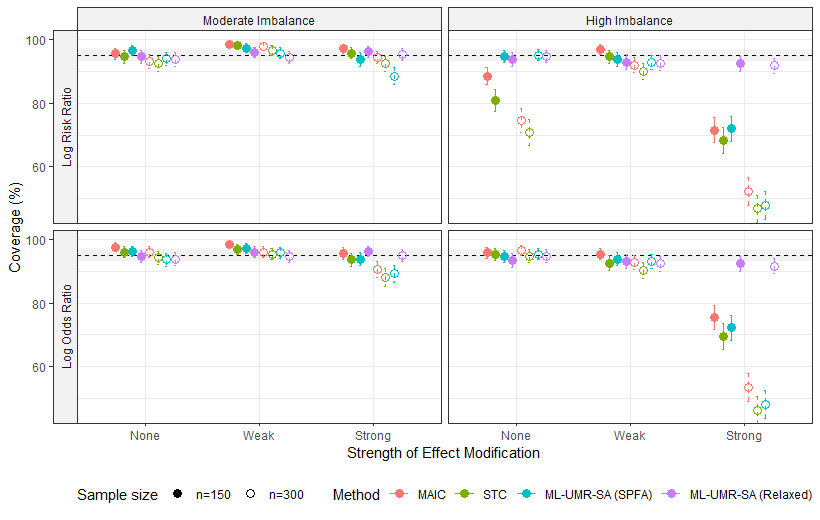}

{\footnotesize
\setlength{\baselineskip}{0.9\baselineskip}
\noindent Abbreviations: MAIC = matching-adjusted indirect comparison; MCSE =
Monte Carlo standard error; ML-NMR = multilevel network meta-regression;
ML-UMR-SA = multilevel unanchored meta-regression for single-arm
indirect comparisons; SPFA = shared prognostic factor assumption; STC =
simulated treatment comparison

Empirical coverage is shown by method and sample size for scenarios
varying effect-modification strength and population imbalance. Error
bars represent ±1.96 × MCSE for each coverage estimate; shading
indicates the Monte Carlo reference band around nominal 95\% coverage
based on 500 replicates. \par
}

\hypertarget{discussion}{%
\section{Discussion}\label{discussion}}

ML-UMR-SA extends the outcome-regression foundation of unanchored STC to a Bayesian framework for synthesizing single-arm evidence from single-arm studies or individual treatment arms of RCTs. The individual-level model in ML-UMR-SA corresponds to the regression typically fitted in STC; however, ML-UMR-SA embeds this model within a joint likelihood that simultaneously incorporates IPD and AgD for all studies and treatments under consideration. Therefore, ML-UMR-SA generalizes pairwise STC to settings with multiple treatments, multiple studies per treatment, and multiple target populations. By formally marginalizing over reconstructed covariate distributions within a multilevel modeling framework, ML-UMR-SA can be viewed as an adaptation of ML-NMR for the unanchored, arm-based setting, modeling absolute treatment-specific outcomes rather than relative treatment effects, retaining its principled integration of IPD and AgD. Although ML-UMR-SA can incorporate individual treatment arms from RCTs as single-arm evidence, doing so does not preserve within-trial randomization or account for correlations between arms from the same trial, potentially introducing bias and underestimating uncertainty. Anchored methods that appropriately account for randomized comparisons should therefore be preferred whenever feasible.

As with all unanchored single-arm indirect comparisons, valid inference relies on strong structural assumptions. In disconnected single-arm evidence, identification (i.e., the ability to separate treatment effects from differences in patient populations and study designs) requires assumptions on the outcome model and cross-study exchangeability. In particular, identification relies on conditional constancy of absolute effects: conditional on the included covariates, absolute outcomes must be comparable across studies. This requires sufficient adjustment for differences in relevant baseline covariates, including prognostic factors and potential effect modifiers, and correct specification of the outcome model. Additional identifying restrictions may be required because the disconnected single-arm data do not identify all treatment-specific covariate effects; in our primary specification, SPFA constrains these effects to be shared across treatments. SPFA is therefore not an additional core population-adjustment assumption, but a restriction imposed for identification under the available data structure. These assumptions are inherently untestable from the observed data and may be difficult to justify in some decision contexts. ML-UMR-SA does not eliminate these challenges; rather, it makes them explicit within a modeling framework that facilitates transportability and sensitivity analysis. In HTA, application of ML-UMR-SA will therefore require transparent justification of identifying restrictions such as SPFA and the assumptions required for transportability, supported by model diagnostics, sensitivity analyses, and reproducible implementation; whether it is used as a primary analysis or to complement existing PAIC approaches will depend on the evidence structure, plausibility of identifying assumptions such as SPFA, target estimand, and practical requirements of the HTA decision context.

From a practical HTA perspective, the distinction between identification
and transport is particularly important because relative treatment
effects are a key input for cost-effectiveness models (CEMs) used to
inform reimbursement decisions. In HTA submissions, unanchored methods,
such as MAIC and STC, are often used to estimate treatment effects in
the comparator study population, which are then implicitly assumed to be
directly transportable to the jurisdiction-specific population of
interest, such as the index study population.\textsuperscript{23} The
conditions required for this direct transportability step to hold are
often not explicitly considered. For ML-UMR-SA, transport to an external
target population requires the conditional outcome model to remain valid
beyond the included studies; thus, measured covariates must be
sufficient to predict absolute outcomes in the target population as well
as within the observed studies. Our simulations demonstrate that
violations of these transportability requirements can lead to biased
estimates in the target population when population differences are
present, even in the absence of effect modification, with bias further
exacerbated under effect modification. These transportability issues
have direct implications for health economic evaluation in HTA, as bias
in relative treatment effects can propagate through CEMs and affect
estimates of incremental costs, health outcomes, and incremental
cost-effectiveness ratios (ICERs). The magnitude and direction of these
downstream effects heavily depend on the structure, inputs, and
assumptions of the particular CEM and may influence reimbursement
decisions, particularly when cost-effectiveness results are close to
decision thresholds. Furthermore, apparent agreement between methods in
the comparator population may give a false sense of robustness, as it
does not ensure validity in the population relevant for decision making.

ML-UMR-SA provides a framework for examining these challenges by
enabling estimation of conditional outcomes and effects, which can be
standardized to obtain marginal treatment outcomes and effects in
decision-relevant populations, rather than relying on implicit direct
transport assumptions. Outcomes and treatment effects can be estimated
at the individual or population level, allowing conditional models and
predictions to inform individual-level simulation models and
standardized marginal outcomes and effects to inform cohort-based CEMs
(e.g., Markov or partitioned survival models). This provides additional
flexibility relative to MAIC, which targets marginal effects, and STC,
which in practice also typically targets marginal effects because
conditional effects are generally infeasible given limited comparator
reporting. The Bayesian framework of ML-UMR-SA also facilitates
probabilistic sensitivity analyses in CEMs, as joint posterior draws of
outcome model parameters can be propagated through standardization and
directly into downstream cost-effectiveness analyses, preserving
parameter uncertainty and correlation. Together, these features allow
analysts to define estimands aligned with the target population used in
health economic evaluations, evaluate the plausibility of identifying
assumptions separately from transport assumptions, and conduct
structured sensitivity analyses that quantify the impact of potential
violations---such as departures from SPFA---on cost-effectiveness
outcomes. In doing so, ML-UMR-SA provides a structured approach for synthesizing single-arm evidence across multiple studies and treatments while making the assumptions underlying decision-relevant estimates explicit.

Future work should focus on methods for assessing and quantifying
violations of the strong assumptions required for unanchored PAICs. The
National Institute for Health and Care Excellence (NICE) Decision
Support Unit (DSU) Technical Support Document (TSD)
18\textsuperscript{17} recommends evaluating residual systematic error
where possible, including out-of-sample comparisons of observed and
predicted outcomes across a set of external studies and in-sample
predictive checks, such as cross-validation, to assess predictive
performance. However, out-of-sample validation is often infeasible in
HTA because relevant external studies in the target population may be
limited and estimation of between-study variance may be unstable.
In-sample checks may underestimate residual error because the IPD study
may be more homogeneous than the target population. Quantitative bias
analysis (QBA) is a particularly promising direction for unanchored
PAICs because it allows analysts to quantify how unmeasured prognostic
factors or effect modifiers would need to behave to materially change
conclusions.\textsuperscript{19} Recent work has proposed QBA for
unanchored STC,\textsuperscript{40} and similar ideas could be adapted
for ML-UMR-SA by incorporating bias parameters or simulated unmeasured
covariates into the outcome regression modeling and standardization
steps. In a Bayesian implementation, uncertainty in these bias
parameters could be propagated directly through posterior treatment
effect estimates and downstream health economic outcomes.

A second area for future work is broader empirical evaluation of the
extensions considered here. The present simulation study evaluates the
core pairwise formulation for binary outcomes. Although a preliminary
proof-of-concept simulation study has evaluated the survival
extension,\textsuperscript{38} broader evaluation of multi-treatment,
multiple-study, and survival settings remains warranted.

A third important area for future work is the development of PAIC methods for partially connected evidence networks, where connected randomized controlled trials and other disconnected evidence sources (e.g., single-arm studies) coexist. Such settings arise frequently in HTA, particularly when new therapies are evaluated in single-arm trials alongside existing randomized comparisons. ML-UMR-SA can incorporate individual arms from RCTs as single-arm evidence, but doing so would not preserve the within-trial randomized comparisons and would therefore fail to leverage the full evidence structure. Extending ML-NMR to jointly synthesize randomized treatment comparisons and disconnected single-arm evidence in partially connected networks would allow both sources of evidence to be incorporated while appropriately accounting for randomization and the additional assumptions required for unanchored comparisons. By bridging otherwise disconnected evidence, such approaches may relax some of the assumptions required by fully unanchored single-arm comparisons. Leveraging randomized comparisons within partially connected networks can help identify relative treatment effects and separate them from study-level differences, while disconnected single-arm evidence can be incorporated through explicit assumptions about baseline risk, prognostic factors, and effect modifiers.

\hypertarget{references}{%
\section{References}\label{references}}

\hspace*{4mm} 1. Lumley T. Network meta-analysis for indirect treatment comparisons.
\emph{Stat Med}. Aug 30 2002;21(16):2313-24. doi:10.1002/sim.1201

2. Dias S, Welton NJ, Sutton AJ, Ades AE. NICE DSU Technical Support
Document 2: A generalised linear modelling framework for pair-wise and
network meta-analysis of randomised controlled trials. 2011;

3. Dias S, Sutton AJ, Ades AE, Welton NJ. Evidence synthesis for
decision making 2: a generalized linear modeling framework for pairwise
and network meta-analysis of randomized controlled trials. \emph{Med
Decis Making}. Jul 2013;33(5):607-17. doi:10.1177/0272989X12458724

4. Dias S, Welton NJ, Sutton AJ, Caldwell DM, Lu G, Ades AE. Evidence
synthesis for decision making 4: inconsistency in networks of evidence
based on randomized controlled trials. \emph{Medical Decision Making}.
2013;33(5):641-656. doi:10.1177/0272989X12455847

5. Hoaglin DC, Hawkins N, Jansen JP, et al. Conducting
indirect-treatment-comparison and network-meta-analysis studies: report
of the ISPOR Task Force on Indirect Treatment Comparisons Good Research
Practices: part 2. \emph{Value Health}. Jun 2011;14(4):429-37.
doi:10.1016/j.jval.2011.01.011

6. Ades A, Welton NJ, Dias S, Phillippo DM, Caldwell DM. Twenty years of
network meta‐analysis: Continuing controversies and recent developments.
\emph{Research synthesis methods}. 2024;15(5):702-727.

7. Phillippo DM, Dias S, Ades AE, Welton NJ. Assessing the performance
of population adjustment methods for anchored indirect comparisons: A
simulation study. \emph{Stat Med}. Dec 30 2020;39(30):4885-4911.
doi:10.1002/sim.8759

8. Phillippo DM, Dias S, Ades AE, Welton NJ. Multilevel network
meta-regression for population-adjusted treatment comparisons.
\emph{Journal of the Royal Statistical Society: Series A}.
2020;183(3):1189-1210. doi:10.1111/rssa.12579

9. Phillippo DM, Dias S, Ades A, Welton NJ. Multilevel network
meta-regression for general likelihoods: synthesis of individual and
aggregate data with applications to survival analysis. \emph{Journal of
the Royal Statistical Society Series A: Statistics in Society}.
2025:qnaf169.

10. Hogervorst MA, Vreman RA, Mantel-Teeuwisse AK, Goettsch WG. Reported
Challenges in Health Technology Assessment of Complex Health
Technologies. \emph{Value Health}. Jun 2022;25(6):992-1001.
doi:10.1016/j.jval.2021.11.1356

11. Tomeczkowski J, Heidbrede T, Eichinger B, et al. Challenges and
Criteria for Single-Arm Trials Leading to an Added Benefit in German
Health Technology Assessments. \emph{Pharmacoeconomics}. Oct
2025;43(10):1223-1233. doi:10.1007/s40273-025-01524-w

12. Bucher HC, Chammartin F. Strengthening health technology assessment
for cancer treatments in Europe by integrating causal inference and
target trial emulation. \emph{Lancet Reg Health Eur}. May
2025;52:101294. doi:10.1016/j.lanepe.2025.101294

13. Patel D, Grimson F, Mihaylova E, et al. Use of External Comparators
for Health Technology Assessment Submissions Based on Single-Arm Trials.
\emph{Value Health}. Aug 2021;24(8):1118-1125.
doi:10.1016/j.jval.2021.01.015

14. Agrawal S, Arora S, Amiri-Kordestani L, et al. Use of Single-Arm
Trials for US Food and Drug Administration Drug Approval in Oncology,
2002-2021. \emph{JAMA Oncol}. Feb 1 2023;9(2):266-272.
doi:10.1001/jamaoncol.2022.5985

15. Pinto CA, Balantac Z, Mt-Isa S, et al. Regulatory benefit-risk
assessment of oncology drugs: A systematic review of FDA and EMA
approvals. \emph{Drug Discov Today}. Oct 2023;28(10):103719.
doi:10.1016/j.drudis.2023.103719

16. Ishak KJ, Proskorovsky I, Benedict A. Simulation and matching-based
approaches for indirect comparison of treatments.
\emph{Pharmacoeconomics}. Jun 2015;33(6):537-49.
doi:10.1007/s40273-015-0271-1

17. Phillippo DM, Ades AE, Dias S, Palmer S, Abrams KR, Welton NJ. NICE
DSU Technical Support Document 18: Methods for population-adjusted
indirect comparisons in submissions to NICE. 2016;

18. Phillippo DM, Ades AE, Dias S, Palmer S, Abrams KR, Welton NJ.
Methods for population-adjusted indirect comparisons in health
technology appraisal. \emph{Medical Decision Making}.
2018;38(2):200-211. doi:10.1177/0272989X17725740

19. Ishak KJ, Chandler C, Liu FF, Klijn S. A Framework for Reliable,
Transparent, and Reproducible Population-Adjusted Indirect Comparisons.
\emph{PharmacoEconomics}. 2025;43(7):691-710.

20. Signorovitch JE, Sikirica V, Erder MH, et al. Matching-adjusted
indirect comparisons: a new tool for timely comparative effectiveness
research. \emph{Value Health}. Sep-Oct 2012;15(6):940-7.
doi:10.1016/j.jval.2012.05.004

21. Signorovitch JE, Wu EQ, Yu AP, et al. Comparative effectiveness
without head-to-head trials: a method for matching-adjusted indirect
comparisons applied to psoriasis treatment with adalimumab or
etanercept. \emph{Pharmacoeconomics}. 2010;28(10):935-45.
doi:10.2165/11538370-000000000-00000

22. Caro JJ, Ishak KJ. No head-to-head trial? simulate the missing arms.
\emph{Pharmacoeconomics}. 2010;28(10):957-67.
doi:10.2165/11537420-000000000-00000

23. Chandler C, Ishak KJ. Reframing population-adjusted indirect
comparisons as a transportability problem: An estimand-based perspective
and implications for health technology assessment. \emph{arXiv preprint
arXiv:260217041}. 2026;doi:10.48550/arXiv.2602.17041

24. Chandler C, Ishak KJ. Anchors Away: Navigating Unanchored Indirect
Comparisons With Multilevel Unanchored Meta-Regression. \emph{Value in
Health}. 2025;28(12)(1):S498.

25. Dias S, Ades AE. Absolute or relative effects? Arm-based synthesis
of trial data. \emph{Res Synth Methods}. Mar 2016;7(1):23-8.
doi:10.1002/jrsm.1184

26. Caflisch RE. Monte Carlo and quasi-Monte Carlo methods. \emph{Acta
Numerica}. 1998;7:1-49.

27. Phillippo D. Calibration of treatment effects in network
meta-analysis using individual patient data {[}PhD thesis{]}. Bristol:
University of Bristol. Available from: University of Bristol repository.
2019;

28. Harari O, Soltanifar M, Cappelleri JC, et al. Network
meta-interpolation: Effect modification adjustment in network
meta-analysis using subgroup analyses. \emph{Res Synth Methods}. Mar
2023;14(2):211-233. doi:10.1002/jrsm.1608

29. Campbell H, Margossian CC, Jansen JP, Gustafson P.
Don\textquotesingle t disregard the data for lack of a likelihood:
Bayesian synthetic likelihood for enhanced multilevel network
meta-regression. \emph{arXiv preprint arXiv:260311019}.
2026;doi:\url{https://doi.org/10.48550/arXiv.2603.11019}

30. Phillippo DM, Dias S, Ades AE, et al. Validating the Assumptions of
Population Adjustment: Application of Multilevel Network Meta-regression
to a Network of Treatments for Plaque Psoriasis. \emph{Med Decis
Making}. Jan 2023;43(1):53-67. doi:10.1177/0272989X221117162

31. Remiro-Azocar A, Heath A, Baio G. Parametric G-computation for
compatible indirect treatment comparisons with limited individual
patient data. \emph{Res Synth Methods}. Nov 2022;13(6):716-744.
doi:10.1002/jrsm.1565

32. Remiro-Azocar A. Target estimands for population-adjusted indirect
comparisons. \emph{Stat Med}. Dec 10 2022;41(28):5558-5569.
doi:10.1002/sim.9413

33. Remiro-Azocar A. Transportability of model-based estimands in
evidence synthesis. \emph{Stat Med}. Sep 30 2024;43(22):4217-4249.
doi:10.1002/sim.10111

34. Remiro-Azócar A, Phillippo DM, Welton NJ, et al. Marginal and
conditional summary measures: transportability and compatibility across
studies. \emph{arXiv preprint arXiv:250721925}. 2025;

35. Sofi-Mahmudi A, Chandler C. mlumr: Multilevel Unanchored
Meta-Regression for Indirect Treatment Comparisons. R package version
0.2.0. 2026. Available from:
\url{https://CRAN.R-project.org/package=mlumr}.

36. Sofi-Mahmudi A, Chandler C. mlumr: Worked Examples and Vignettes.
2026. Available from: \url{https://choxos.github.io/mlumr/articles/}.

37. Guyot P, Ades AE, Ouwens MJ, Welton NJ. Enhanced secondary analysis
of survival data: reconstructing the data from published Kaplan-Meier
survival curves. \emph{BMC Med Res Methodol}. Feb 1 2012;12:9.
doi:10.1186/1471-2288-12-9

38. Chandler C, Ishak J. Surviving Unanchored Indirect Comparisons: An
Extension of Multilevel Unanchored Meta-Regression (ML-UMR) for Survival
Analyses. \emph{Value in Health}. 2026;29(S6):S324.

39. Chandler C, Sofi-Mahmudi A. Simulation code for ``Anchors Away:
Navigating Single-Arm Indirect Comparisons with Multilevel Unanchored
Meta-Regression (ML-UMR-SA).'' GitHub. 2026. Available from:
\url{https://github.com/choxos/mlumr/tree/simulation}.

40. Ren S, Ren S, Welton NJ, Strong M. Quantitative bias analysis for
unmeasured confounding in unanchored population-adjusted indirect
comparisons. \emph{Res Synth Methods}. May 2025;16(3):509-527.
doi:10.1017/rsm.2025.13

\begin{Backmatter}

\vspace{6mm}

\paragraph*{Acknowledgments}
The authors are especially grateful to Ahmad Sofi-Mahmudi for his initiative, contributions, and continued effort in developing the \emph{mlumr} R package that supported the implementation of this work. The authors also thank Ruth Sharf for editorial support on this manuscript. The authors also used OpenAI's ChatGPT 5.5 to assist with proofreading and copyediting of manuscript text.

\paragraph*{Funding Statement}
This study and article were funded by Thermo Fisher Scientific. The authors are employees of Thermo Fisher Scientific.

\paragraph*{Competing Interests}
The authors declare none.

\paragraph*{Data Availability Statement}
A simulation study is presented in this article. All simulations were conducted using the \emph{mlumr} R package, with the full code available on GitHub:  https://github.com/choxos/mlumr/tree/simulation.

\paragraph*{Author Contributions}
Conor O. Chandler: Conceptualization, methodology, writing -- original draft, writing --
reviewing and editing, visualization, formal analysis; K. Jack Ishak:
methodology, writing -- reviewing and editing.

\end{Backmatter}

\newpage

\noindent \textbf{Supplementary Material for ``Anchors Away: Navigating Single-Arm Indirect Comparisons with Multilevel Unanchored Meta-Regression (ML-UMR-SA)''}

\appendix

\section{Identification Assumptions for Unanchored PAICs}

In fully disconnected single-arm evidence, each study evaluates a single
treatment. Treatment effects are therefore not separately identifiable
from study, population, and design effects without additional structural
assumptions. Consequently, multilevel unanchored meta-regression for
single-arm comparisons (ML-UMR-SA) requires assumptions that are
stronger than those needed for anchored indirect comparisons.

The assumptions in A.1--A.5 are common to unanchored population-adjusted
indirect comparison (PAIC) methods, including matching-adjusted indirect
comparison (MAIC) and simulated treatment comparison (STC). ML-UMR-SA
does not remove these requirements. Instead, it embeds these assumptions
within a multilevel likelihood framework that generalizes beyond
pairwise comparisons and supports transport of marginal and conditional
effects to decision-relevant populations. Additionally, the disconnected
single-arm data structure may require identifying restrictions on the
covariate--outcome model when treatment-specific covariate effects are
not identified from the available data (A.6). Within ML-UMR-SA, this
issue arises explicitly through the joint modeling of treatment-specific
covariate effects and the restrictions required to identify them when
covariate--outcome information is limited. These restrictions are not
additional core assumptions beyond A.1--A.5, but model-specific
identifying restrictions imposed under limited information.

\subsection{Conditional Exchangeability of Potential Outcomes Across
Studies (No Unmeasured
Confounding)}

Unanchored PAICs assume conditional exchangeability of potential
outcomes across studies, meaning that, conditional on baseline
covariates \(X\), study membership \(J\) is independent of potential
outcomes \(Y^{(k)}\) under each treatment \(k\):

\begin{align}
Y^{(k)} \perp\!\!\!\perp J \mid X,\quad \forall k. \notag
\end{align}

Equivalently, all study or population differences that influence
outcomes are assumed to be captured by \(X\). In unanchored comparisons,
this includes all prognostic factors and all treatment effect modifiers
on the chosen modeling scale. National Institute for Health and Care
Excellence (NICE) Decision Support Unit (DSU) Technical Support Document
(TSD) 18{[}1{]} refers to this requirement as \emph{``conditional
constancy of absolute effects.''} In the causal inference literature, a
closely related assumption is Strongly Ignorable Treatment Assignment
(SITA), under which treatment (here, study membership) is independent of
potential outcomes conditional on measured covariates.{[}2{]}

\hypertarget{a.2-positivity-overlap-and-extrapolation}{%
\subsection{Positivity / Overlap and
Extrapolation}\label{a.2-positivity-overlap-and-extrapolation}}

For any target population \(\widetilde{P}\) to which effects are
transported, there must be sufficient covariate overlap between
\(\widetilde{P}\) and the study populations informing each treatment
model. Formally, for covariate values \(x\) with positive density in the
target population,

\[
f_{\widetilde{P}}(x) > 0 \quad\Rightarrow\quad f_{j,k}(x) > 0.
\]

for studies \(j\) contributing information about treatment \(k\).{[}2,
3{]}

When overlap is limited, estimation relies on extrapolation beyond the
observed data.{[}4{]}

\hypertarget{a.3-stable-unit-treatment-value-assumption-sutva-consistency-and-no-interference}{%
\subsection{Stable Unit Treatment Value Assumption (SUTVA):
Consistency and No
Interference}\label{a.3-stable-unit-treatment-value-assumption-sutva-consistency-and-no-interference}}

All PAICs assume consistency, meaning that the observed outcome equals
the potential outcome under the treatment actually received:

\[Y = Y^{(T)}.\]

This requires that treatment \(k\) is well-defined and corresponds to a
single, coherent intervention across studies.{[}3{]} In particular,
there must be no systematically different ``versions'' of treatment
\(k\)--- arising from differences in study protocols, background care,
outcome definitions, or implementation --- that induce distinct
potential outcomes, unless such differences are captured by measured
covariates \(X\) or explicitly modeled (e.g., through study-level
effects). PAICs also assume no interference: one individual's potential
outcomes are unaffected by the treatment assignments of other
individuals.{[}3{]}

\hypertarget{a.4-correct-specification-of-the-outcome-regression-model}{%
\subsection{Correct Specification of the Outcome Regression
Model}\label{a.4-correct-specification-of-the-outcome-regression-model}}

Valid inference requires correct specification of the conditional
outcome model,

\[\mathbb{E}\left\{ Y\mid X,k \right\},\]

on the chosen modeling scale. This includes appropriate functional
forms, prognostic effects, and treatment--covariate interactions
representing effect modification.{[}3, 5-8{]}

Both STC and ML-UMR-SA depend on a correctly specified conditional outcome
model.{[}4, 5{]}

\hypertarget{a.5-valid-reconstruction-of-the-joint-covariate-distribution-in-agd-studies}{%
\subsection{Valid Reconstruction of the Joint Covariate Distribution
in AgD
Studies}\label{a.5-valid-reconstruction-of-the-joint-covariate-distribution-in-agd-studies}}

Marginal summaries of baseline characteristics in AgD studies are
generally insufficient to calculate adjusted outcomes and effects
because marginal estimands depend on the joint covariate distribution.
STC and ML-UMR-SA, therefore, require an assumed or reconstructed joint
covariate distribution for AgD studies, and valid estimation depends on
how closely this distribution approximates the true covariate
distribution in the AgD population. In weighting-based approaches such
as MAIC, the analogous requirement is that the weights applied to the
individual patient data (IPD) population successfully balance the
covariate functions needed to identify the target estimand. Matching
only reported marginal summaries may be insufficient when the outcome
model depends on nonlinear terms, interactions, or unreported aspects of
the joint covariate distribution.

\hypertarget{a.6-additional-identification-requirements-for-ml-umr-spfa-or-subgroup-identifying-information}{%
\subsection{Identifying Restrictions under Limited Single-Arm
Evidence}\label{a.6-additional-identification-requirements-for-ml-umr-spfa-or-subgroup-identifying-information}}

Within ML-UMR-SA, when one or more comparator studies contribute only
aggregate outcomes with limited covariate--outcome information,
treatment-specific covariate effects may not be separately identified.
Identification then requires additional information or restrictions
relating covariate effects across treatments. A primary identifying
restriction considered in ML-UMR-SA is the shared prognostic factor
assumption (SPFA):

\[\beta_{1} = \beta_{2} = \cdots = \beta_{K},\]

which constrains covariate effects to be common across treatments on the
model's linear predictor scale.{[}9, 10{]} SPFA is not an additional
core unanchored PAIC assumption beyond A.1--A.5; rather, it is an
identifying restriction imposed because the available single-arm
evidence does not generally identify treatment-specific covariate
effects.

Under SPFA, the covariate--outcome relationship estimated from IPD for
one treatment informs the prognostic component of the outcome model for
treatments observed only through AgD. This constraint links treatments
through shared regression parameters, enabling separation of
treatment-specific intercepts from covariate effects and thereby
permitting identification of treatment contrasts under the assumptions
in A.1--A.5. This role is analogous to the SEMA that may be imposed for
identification in anchored ML-NMR settings.

If SPFA does not adequately represent the true treatment-specific
covariate--outcome relationships, the resulting outcome model is
misspecified, relating violations of SPFA to the conditional
exchangeability and outcome-model specification requirements described
in A.1 and A.4. When SPFA is relaxed (i.e., \(\beta_{k}\) differs across
treatments), identification typically requires richer aggregate
information (e.g., joint subgroup outcomes) or sensitivity analyses
using informative priors on treatment-specific deviations.

\hypertarget{a.7-contrast-with-anchored-indirect-comparisons}{%
\subsection{Contrast with Anchored Indirect
Comparisons}\label{a.7-contrast-with-anchored-indirect-comparisons}}

Anchored methods (e.g., network meta-analysis {[}NMA{]} and multilevel
network meta-regression {[}ML-NMR{]}) compare treatment contrasts
estimated within randomized controlled trials. Randomization supports
conditional exchangeability of treatment assignments within each trial.
Across studies, anchored methods assume that relative treatment effects
are conditionally exchangeable (or constant) given effect modifiers, an
assumption referred to in NICE DSU TSD 18{[}1{]} as \emph{``conditional
constancy of relative effects.''}

In contrast, unanchored comparisons lack within-trial randomization
linking treatments. They, therefore, require conditional exchangeability
of absolute potential outcomes across studies (A.1), which entails
adjustment for all prognostic factors as well as effect modifiers, and
additional identifying structure such as SPFA (A.6). Unlike anchored
approaches such as ML-NMR, which can allow study-specific intercepts for
residual baseline risk differences, unanchored PAICs typically require a
common intercept (i.e., conditional baseline risk) across study
populations because study-level effects are not separately identifiable
from treatment effects in disconnected single-arm evidence. This matters
particularly for absolute outcomes and marginal effects, where
transportability may depend on baseline risk as well as effect
modification.

\hypertarget{appendix-b.-pooling-approaches-for-synthesizing-multiple-studies-per-treatment}{%
\section{Pooling Approaches for Synthesizing Multiple
Studies per
Treatment}\label{appendix-b.-pooling-approaches-for-synthesizing-multiple-studies-per-treatment}}

Pooling decisions should be guided by the number of studies per
treatment, the degree of observed heterogeneity, and the plausibility of
exchangeability of baseline risk across studies. The pooling strategies
for ML-UMR-SA correspond to standard meta-analytic approaches but can be
implemented within a unified Bayesian framework for ML-UMR-SA.

\hypertarget{no-pooling-fixed-study-effects}{%
\subsection{No Pooling (Fixed Study
Effects)}\label{no-pooling-fixed-study-effects}}

Let \(j = 1,\ldots,J\) index studies and \(k(j) \in \{ 1,\ldots,K\}\)
denote the treatment administered in study \(j\). Under no pooling of
baseline outcome parameters, each study is assigned its own intercept.
Let \(\alpha_{j}\) denote a study-specific intercept capturing the
study-specific baseline outcome level conditional on the modeled
covariates. The linear predictor becomes

\[\eta_{ij} = \alpha_{j} + x_{ij}^{\top}\beta_{k(j)},\]

with \(Y_{ij} \mid x_{ij} \sim \pi(g^{- 1}(\eta_{ij}))\).

For AgD, the study-specific marginal mean is

\[
\theta_{\bullet j}
=
\int
g^{-1}\!\left(\alpha_j + x^{\top}\beta_{k(j)}\right)
\, f_j(x)\, dx.
\]

The full likelihood retains the product form

\[
\mathcal{L}
=
\prod_{j \in \mathcal{S}_{\mathrm{IPD}}}
\prod_{i=1}^{n_j}
\pi_{\mathrm{IPD}}\!\left(y_{ij}\mid\theta_{ij}\right)
\,
\prod_{j \in \mathcal{S}_{\mathrm{AgD}}}
\pi_{\mathrm{AgD}}\!\left(y_{\bullet j}\mid\theta_{\bullet j}\right).
\]

Let \(\mathcal{S}_{IPD}\) and \(\mathcal{S}_{AgD}\) denote the set of
IPD and AgD studies.

For a target population with covariate distribution
\(f_{\widetilde{P}}(x)\), define standardized study-level means as

\[
\mu_{j,\widetilde{P}}
=
\int
g^{-1}\!\left(\alpha_j + x^{\top}\beta_{k(j)}\right)
\,f_{\widetilde{P}}(x)\,dx.
\]

A pooled marginal contrast for A vs. B on the linear predictor scale is

\[
\Delta_{A,B}(\widetilde{P})
=
\sum_{j \in \mathcal{S}_A}
\sum_{j' \in \mathcal{S}_B}
w_{jj'}
\bigl[
h(\mu_{j,\widetilde{P}})
-
h(\mu_{j',\widetilde{P}})
\bigr].
\]

where \(h( \cdot )\) denotes the chosen effect scale and
\(\sum w_{jj^{'}} = 1\). In the Bayesian implementation, the weighted
average for the pooled marginal contrast for A vs. B is evaluated at
each posterior draw. Weights may be defined based on sample size,
precision, or other decision-relevant criteria.

This may be viewed as a two-stage approach: study-specific outcome
models are first estimated and standardized to a common target
population, after which the resulting predictions or contrasts are
combined. Alternatively, study-specific comparisons may be retained
separately rather than pooled. Neither approach removes the conditional
constancy of absolute effects assumption required to compare absolute
outcomes across disconnected single-arm studies.

\hypertarget{complete-pooling-common-baseline-within-treatment}{%
\subsection{Complete Pooling (Common Baseline Within
Treatment)}\label{complete-pooling-common-baseline-within-treatment}}

Under complete pooling, studies evaluating the same treatment are
assumed to share a common baseline risk after covariate adjustment. That
is, intercepts are indexed by treatment rather than study. The linear
predictor becomes

\[\eta_{ij} = \alpha_{k(j)} + x_{ij}^{\top}\beta_{k(j)}.\]

Here, all studies evaluating treatment \(k\) contribute to the
estimation of a single treatment-specific intercept \(\alpha_{k}\).
Differences between studies are attributed solely to observed covariates
and sampling variability.

Complete pooling defines a treatment-level effect directly through the
shared intercept \(\alpha_{k}\). However, this approach imposes an
additional restriction of no residual between-study heterogeneity in
baseline outcome parameters beyond that explained by observed
covariates. This restriction may only be plausible in settings where
studies evaluating the same treatment can reasonably be regarded as
close replications (e.g., studies conducted under a common protocol or
within the same development program) and is unlikely to be plausible in
many disconnected single-arm applications. Importantly, alternative
pooling strategies relax how residual heterogeneity in baseline
parameters is modeled but do not remove the conditional constancy of
absolute effects assumption required to compare outcomes across
disconnected single-arm studies. This assumption is itself strong, may
be difficult to justify in practice, and violations may result in biased
or overprecise inference.

\hypertarget{partial-pooling-of-study-level-baselines}{%
\subsection{Partial Pooling of Study-level
Baselines}\label{partial-pooling-of-study-level-baselines}}

When a sufficiently large number of studies are available, between-study
heterogeneity in baseline risk may be modeled hierarchically by assuming
exchangeability of baseline risk across studies (i.e., study-level
random intercepts). For two treatments \(A\) and \(B\), one convenient
parameterization is

\[
g(\theta_{ijk})
=
\mu_j
+
\delta\,\mathds{1}\{k(j)=B\}
+
x_{ij}^{\top}\beta.
\]

with

\[\begin{matrix}
 & \mu_{j} \sim \mathcal{N}(\mu_{0},\tau_{\mu}^{2}). & & \\
\end{matrix}\]

For \(K > 2\), treatment indicators may be included analogously. Here,
\(\mu_{j}\) captures residual study-level baseline variation after
covariate adjustment, while \(\delta\) parameterizes a common conditional
treatment effect on the linear predictor scale. This formulation induces
partial pooling of baseline risk across studies while preserving a
common conditional treatment effect.

In practice, estimation of \(\tau_{\mu}^{2}\) requires adequate
replication (e.g., ideally at least 3--4 studies per treatment). With
sparse evidence---particularly in disconnected single-arm settings where
each study contributes only a single aggregate endpoint---hierarchical
baseline models may be weakly identified. In such cases, posterior
inference can be highly sensitive to prior assumptions on
\(\tau_{\mu}\), and treatment contrasts may be driven largely by
exchangeability assumptions rather than the observed data. That is,
because treatment is confounded with study in disconnected single-arm
evidence, separation of the treatment effect \(\delta\) from study-level
baseline heterogeneity relies on the assumed exchangeability
distribution for the study-specific baselines. We therefore present
fixed study effects as the primary specification and recommend
hierarchical pooling as a sensitivity analysis when sufficient
replication across studies exists.

If a sufficiently rich set of data is available, alternative partial
pooling approaches could be specified and explored. However, in
practice, such data may not be available to support such model
specification.

\hypertarget{appendix-c.-ml-umr-simulation-study-details}{%
\section{ML-UMR-SA Simulation Study
Details}\label{appendix-c.-ml-umr-simulation-study-details}}

\hypertarget{c.1-ademp-summary}{%
\subsection{ADEMP Summary}\label{c.1-ademp-summary}}

The simulation study is summarized using the ADEMP framework (Aims,
Data-generating mechanisms, Estimands, Methods, and Performance
measures), a structured approach for reporting simulation
studies.{[}11{]}

\setlength{\tabcolsep}{4pt}
{\footnotesize
\begin{longtable}{@{}L{0.18\linewidth} L{0.78\linewidth}@{}}
\toprule
\textbf{Component} & \textbf{Specification} \\
\midrule
\endfirsthead

\toprule
\textbf{Component} & \textbf{Specification} \\
\midrule
\endhead

\midrule
\multicolumn{2}{r}{\footnotesize Continued on next page} \\
\midrule
\endfoot

\bottomrule
\endlastfoot

Aims &
Evaluate performance of ML-UMR-SA versus unanchored STC, unanchored MAIC, and naïve comparisons; assess sensitivity to SPFA violations, population imbalance, and misspecification of covariate dependence. \\

Data-generating mechanisms &
Two independent single-arm studies (A: IPD; B: AgD). Binary outcome generated from logistic regression with treatment-specific intercepts and covariate effects. Effect modification induced by allowing treatment-specific covariate coefficients.

\smallskip

Factorial design varying EM strength, imbalance, and covariate dependence.

\smallskip

500 Monte Carlo replications per scenario. \\

Estimands &
Marginal treatment effects (A vs B) in both the comparator population and index population, on the log-odds ratio and log risk ratio scales. \\

Methods &
ML-UMR-SA (SPFA), ML-UMR-SA (relaxed SPFA using subgroup AgD), unanchored STC (marginalized g-computation), unanchored MAIC (entropy balancing), naïve comparison.

\smallskip

For STC and MAIC, adjusted relative effects are estimated in the comparator population, and the same relative effects are then assumed for the index population, as is commonly done in applied practice. This deliberate assumption helps illustrate potential transport bias with MAIC and STC. \\

Performance measures &
Bias, absolute bias, relative bias (\%), empirical SE, RMSE, 95\% coverage, interval width, Monte Carlo SEs. \\

\end{longtable}
}

{\footnotesize
\setlength{\baselineskip}{0.9\baselineskip}
\noindent Abbreviations: AgD = aggregate data; EM = effect modification; IPD =
individual patient data; MAIC = matching-adjusted indirect comparison;
ML-UMR-SA = multilevel unanchored meta-regression; RMSE = root mean squared
error; SE = standard error; SPFA = shared prognostic factor assumption;
STC = simulated treatment comparison. \par
}

\hypertarget{c.2-aims}{%
\subsection{Aims}\label{c.2-aims}}

This simulation study evaluates the finite-sample performance of ML-UMR-SA
for unanchored indirect comparisons using disconnected single-arm
evidence, where IPD are available for the index treatment (A) and only
AgD for the comparator (B).

The study is designed to explicitly assess the role of identifying
assumptions---particularly the SPFA---and to contrast ML-UMR-SA with
commonly used PAIC methods.

\hypertarget{c.3-data-generating-model}{%
\subsection{Data-Generating Model}\label{c.3-data-generating-model}}

\hypertarget{c.3.1-study-structure-and-data-availability}{%
\subsubsection{Study Structure and Data
Availability}\label{c.3.1-study-structure-and-data-availability}}

Each simulation replicate consists of two independent single-arm
studies: an index study (A) with IPD and a comparator study (B) with AgD
only. The studies are fully disconnected (i.e., there is no shared
comparator), so treatment effects are not directly identifiable without
additional structural assumptions.

For the index study (A), IPD are assumed to be available. For the
comparator study (B), however, only AgD are available. Under the shared
prognostic factor assumption (SPFA), only overall outcome information
(e.g., event counts) and marginal covariate summaries are available.
When SPFA is relaxed, richer AgD are provided in the form of
subgroup-level outcomes, where the comparator population is partitioned
into four non-overlapping subgroups defined by interactions between
\(X_{1}\) and \(X_{2}\).

\hypertarget{c.3.2-covariates}{%
\subsubsection{Covariates}\label{c.3.2-covariates}}

Two binary baseline covariates, \(X = (X_{1},X_{2})\), were generated in
each study using a Gaussian copula with calibrated correlation. The
simulation varied both marginal covariate prevalence, to induce
population imbalance between the index and comparator populations, and
the covariate dependence structure, to evaluate sensitivity to
misspecification of the joint covariate distribution. The specific
prevalence and correlation values used in each scenario are summarized
in Table C.1.

\hypertarget{c.3.3-outcome-model}{%
\subsubsection{Outcome Model}\label{c.3.3-outcome-model}}

Outcomes were generated from

\[\text{logit}\{ P(Y = 1 \mid X,k)\} = \alpha_{k} + X^{\top}\beta_{k},
\]

with \(\alpha_{A} = 1.0\), \(\alpha_{B} = 0.25\), and
\(\beta_{A} = ( - 1.0, - 2.0)\). Effect modification was introduced
through the coefficient for \(X_{2}\) under treatment B:

\[\beta_{B} = \beta_{A} + \delta.\]

The three effect-modification settings were evaluated (Table C.1).

\hypertarget{c.3.4-scenario-grid}{%
\subsubsection{Scenario Grid}\label{c.3.4-scenario-grid}}

Four factors were varied in a factorial design, yielding a total of
\(3 \times 2 \times 2 \times 2 = 24\) scenarios. Each scenario was
evaluated using 500 Monte Carlo replications. Model estimation was
performed using Bayesian MCMC with 3 chains and 2,000 iterations per
chain.

\vspace{3mm}
\noindent \textbf{Table C.1} Scenario factors and levels

\setlength{\tabcolsep}{4pt}
{\footnotesize
\renewcommand{\arraystretch}{1.15}
\begin{longtable}{@{}L{0.17\linewidth} L{0.37\linewidth} L{0.40\linewidth}@{}}
\toprule
\textbf{Factor} & \textbf{Levels} & \textbf{Description} \\
\midrule
\endfirsthead

\toprule
\textbf{Factor} & \textbf{Levels} & \textbf{Description} \\
\midrule
\endhead

\midrule
\multicolumn{3}{r}{\footnotesize Continued on next page} \\
\midrule
\endfoot

\bottomrule
\endlastfoot

\textbf{EM} &
None (SPFA): \(\beta_{B,2}=-2.0\)

\smallskip
Weak (12.5\% change): \(\beta_{B,2}=-1.75\)

\smallskip
Strong (50\% change): \(\beta_{B,2}=-1.0\) &
Variation in treatment-specific covariate effect for \(X_2\); \(\beta_{B,1}=-1.0\) is held constant. \\

\textbf{Population imbalance} &
Moderate: \(P(X_j=1)=0.60\) (index), \(0.40\) (comparator)

\smallskip
High: \(P(X_j=1)=0.80\) (index), \(0.20\) (comparator) &
Differences in marginal covariate prevalences between populations. \\

\textbf{Covariate dependence} &
Correct: \(\rho=0.50\) (both populations)

\smallskip
Misspecified: \(\rho=0.50\) (index), \(0.25\) (comparator) &
Correlation between \(X_1\) and \(X_2\). \\

\textbf{Sample size} &
\(n_A=n_B=150, 300\) &
Number of individuals per study. \\

\end{longtable}
}

{\footnotesize
\setlength{\baselineskip}{0.9\baselineskip}
\noindent Abbreviations: EM = effect modification; SPFA = shared prognostic factor
assumption \par
}

\hypertarget{c.4-estimands}{%
\subsection{Estimands}\label{c.4-estimands}}

For a target population \(P \in \{\text{index},\text{comparator}\}\),
the marginal outcome under treatment \(k \in \{ A,B\}\) is defined as:

\[\mu_{k,P} = \mathbb{E}_{X \sim f_{P}}\left\lbrack \text{logit}^{- 1}\left( \alpha_{k}+X^{\top}\beta_{k} \right) \right\rbrack.\]

The target estimands are marginal treatment effects evaluated in
population \(P\) on the log-odds ratio and log risk ratio scales:

\begin{align*}
\Delta_{P}^{\mathrm{LOR}}
&=
\operatorname{logit}(\mu_{A,P})
-
\operatorname{logit}(\mu_{B,P}),
\\
\Delta_{P}^{\mathrm{RR}}
&=
\log(\mu_{A,P})
-
\log(\mu_{B,P}).
\end{align*}

True marginal treatment effects were computed for each scenario using
large simulated populations (N=10,000,000) to approximate the
data-generating distributions.

\hypertarget{c.5-methods}{%
\subsection{Methods}\label{c.5-methods}}

\hypertarget{c.5.1-ml-umr-spfa}{%
\subsubsection{ML-UMR-SA (SPFA)}\label{c.5.1-ml-umr-spfa}}

ML-UMR-SA is implemented using a logistic outcome regression model. Under
the SPFA, covariate effects are assumed to be identical across
treatments, such that:

\[\text{logit}\left\{ P\left( Y = 1\mid X,k \right) \right\} = \alpha_{k} + X^{\top}\beta,\ \ \ \ \text{with\ }\beta_{A} = \beta_{B} = \beta.\]

For the index study, outcomes are modeled at the individual level as:

\[
Y_i \sim \operatorname{Bernoulli}(p_i),
\qquad
\operatorname{logit}(p_i)
=
\alpha_A + X_i^{\top}\beta.
\]

For the comparator study, only AgD are observed. Let \(N_{B}\) and
\(R_{B}\) denote the total sample size and number of events. The
marginal event probability in the comparator population is:

\[\mu_{B,\text{comp}} = \mathbb{E}_{X \sim f_{\text{comp}}}\left\lbrack \text{logit}^{- 1}(\alpha_{B} + X^{\top}\beta) \right\rbrack,\]

and the likelihood contribution is:

\[R_{B} \sim \text{Binomial}(N_{B},\mu_{B,\text{comp}}).\]

This formulation uses a one-parameter Binomial likelihood, where the
event probability \(\mu_{B,\text{comp}}\) represents the marginal mean
of individual event probabilities in the comparator population, i.e.
\(\mu_{B,\text{comp}} = E_{X}\lbrack p(X)\rbrack\). This implicitly
assumes that the aggregate outcomes are fully characterized by this
mean, without additional dispersion beyond that induced by the Binomial
model. Alternative formulations have been proposed that relax this
assumption. In particular, Phillippo et al.{[}12{]} describe a
two-parameter aggregate likelihood that allows additional flexibility
(e.g., to account for overdispersion or heterogeneity in the
distribution of individual risks). While such extensions were not
explored here, they are implemented in the \emph{multinma} and
\emph{mlumr} R packages.

Because individual-level outcomes and covariates are not observed for
the comparator population, the marginal outcome is approximated using
numerical integration:

\[\mu_{B,\text{comp}} \approx \frac{1}{Q}\sum_{q = 1}^{Q}\text{logit}^{- 1}(\alpha_{B} + X_{q}^{\top}\beta),\]

where \(X_{q}\) are integration points drawn from a parametric
approximation to \(f_{\text{comp}}(X)\). \(Q = 512\) integration points
were used.

Posterior draws of \(\mu_{k,P}\) are obtained for
\(P \in \{\text{index},\ \text{comparator}\}\), from which treatment
effects $\Delta_{P}^{\mathrm{LOR}}$ and $\Delta_{P}^{\mathrm{RR}}$ are computed.
Inference is fully Bayesian, with point estimates given by posterior
means and uncertainty summarized using 95\% credible intervals.

\hypertarget{c.5.2-ml-umr-relaxed-spfa}{%
\subsubsection{ML-UMR-SA (Relaxed
SPFA)}\label{c.5.2-ml-umr-relaxed-spfa}}

ML-UMR-SA was extended to relax SPFA by allowing treatment-specific
covariate effects:

\[\text{logit}\left\{ P\left( Y = 1\mid X,k \right) \right\} = \alpha_{k} + X^{\top}\beta_{k},\ \ \ \ \ \ \text{with\ }\beta_{A} \neq \beta_{B}.\]

As in the SPFA specification, the index study contributes an
individual-level likelihood based on observed outcomes:

\[Y_{i} \sim \text{Bernoulli}\left( p_{i} \right),\ \ \ \ \text{logit}(p_{i}) = \alpha_{A} + X_{i}^{\top}\beta_{A}.\]

For the comparator study, identification of \(\beta_{B}\) requires
richer aggregate information than overall event counts. This is achieved
by incorporating subgroup-level AgD. Let \(s = 1,\ldots,S\) index
mutually exclusive subgroups defined by baseline covariates (e.g., joint
strata of \(X_{1}\) and \(X_{2}\)). For each subgroup, let \(N_{B,s}\)
and \(R_{B,s}\) denote the subgroup sample size and number of events.
The likelihood contribution is:

\[R_{B,s} \sim \text{Binomial}(N_{B,s},\mu_{B,s}),\]

where the subgroup-specific marginal outcome is:

\[\mu_{B,s} = \mathbb{E}_{X \sim f_{s}}\left\lbrack \text{logit}^{- 1}(\alpha_{B} + X^{\top}\beta_{B}) \right\rbrack,\]

estimated using numerical integration with \(Q = 128\) integration
points.

\hypertarget{c.5.3-stc-g-computation}{%
\subsubsection{STC
(G-Computation)}\label{c.5.3-stc-g-computation}}

STC is implemented using outcome regression with marginalization
(g-computation).{[}5, 7, 13-15{]} A logistic regression model is first
fitted to the IPD for the index treatment:

\[\text{logit}\{ P(Y = 1 \mid X)\} = \gamma_{0} + X^{\top}\gamma,\]

which estimates the conditional outcome model under treatment \(A\).

This model is then used to predict outcomes in the comparator population
by averaging over its covariate distribution:

\[{\widehat{\mu}}_{A,\text{comp}} = \frac{1}{Q}\sum_{q = 1}^{Q}\text{logit}^{- 1}(\gamma_{0} + X_{q}^{\top}\gamma),\]

where \(X_{q}\) are integration points representing the covariate
distribution \(f_{\text{comp}}(X)\). \(Q = 512\) integration points were
used.

The marginal outcome under treatment \(B\)in the comparator population
is obtained directly from AgD:

\[{\widehat{\mu}}_{B,\text{comp}} = \frac{R_{B}}{N_{B}}.\]

Treatment effects on the log-odds ratio and log risk ratio scales are
obtained by applying the estimand definitions in Section C.4 to the
estimated marginal outcomes.

To obtain effects in the index population, STC assumes that relative
treatment effects are directly transportable (i.e., invariant) across
populations, so that:

\[\Delta_{\text{index}} = \Delta_{\text{comparator}}.\]

Uncertainty is estimated using the delta method, combining model-based
uncertainty from the fitted regression with binomial variance for the
comparator outcome.

\hypertarget{c.5.4-maic}{%
\subsubsection{MAIC}\label{c.5.4-maic}}

MAIC is implemented using weighting to align the covariate distribution
of the index study with that of the comparator population.{[}16-18{]}
Weights \(w_{i}\) are estimated using entropy balancing so that weighted
covariate means match the reported aggregate summaries:

\[
\sum_i w_i X_i = \bar{X}_{\text{comp}}.
\]

The weights take the form:

\[w_{i} = \exp(X_{i}^{\top}\alpha),\]

where \(\alpha\) is optimized to satisfy the moment constraints.

The marginal outcome under treatment \(A\) in the comparator population
is estimated using a weighted mean:

\[{\widehat{\mu}}_{A,\text{comp}} = \frac{\sum_{i}^{}w_{i}Y_{i}}{\sum_{i}^{}w_{i}}.\]

The corresponding comparator outcome is obtained directly from AgD:

\[{\widehat{\mu}}_{B,\text{comp}} = \frac{R_{B}}{N_{B}}.\]

Treatment effects are obtained by applying the estimand definitions in
Section C.4 to these estimated marginal outcomes. Uncertainty is
estimated using robust (sandwich) standard errors (HC3) for the weighted
estimator \({\widehat{\mu}}_{A,\text{comp}}\), combined with
conventional binomial variance for the comparator outcome.

To obtain effects in the index population, MAIC assumes that relative
treatment effects are directly transportable (i.e., invariant across
populations):

\[\Delta_{\text{index}} = \Delta_{\text{comparator}}.\]

\hypertarget{c.5.5-nauxefve-comparison}{%
\subsubsection{Naïve Comparison}\label{c.5.5-nauxefve-comparison}}

The naïve estimator compares observed outcomes directly. Variance is
estimated using standard large-sample approximations based on binomial
variability.

\hypertarget{c.6-performance-measures}{%
\subsection{Performance Measures}\label{c.6-performance-measures}}

Performance was primarily assessed using bias and coverage of 95\%
intervals, with absolute bias, relative bias, empirical standard error,
RMSE, interval width, and Monte Carlo standard errors reported as
supplementary metrics.

The true marginal outcomes and treatment effects (i.e., target
estimands) for each scenario and additional simulation results are
available in the corresponding .csv files in the Supplementary
Materials.

\hypertarget{c.7-additional-simulation-results}{%
\subsection{Sensitivity Analysis: Effect of Sample Size}\label{c.7-additional-simulation-results}}

The following sensitivity analyses considered a larger sample size of n=300 per arm.

\vspace{3mm}
\noindent \textbf{Fig. C.1} Bias for Log Risk Ratio (n=300)

\includegraphics[scale=0.5]{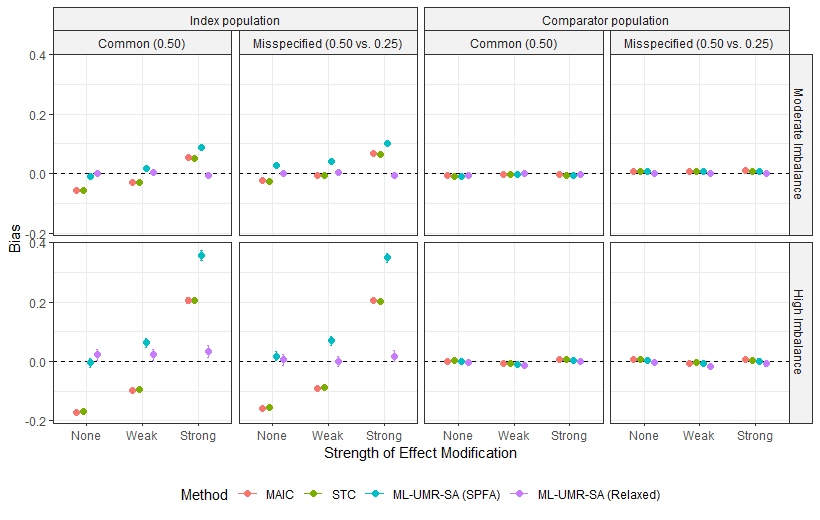}

{\footnotesize
\setlength{\baselineskip}{0.9\baselineskip}
\noindent Abbreviations: EM = effect modification; MAIC = matching-adjusted
indirect comparison; ML-UMR-SA = multilevel unanchored meta-regression;
SPFA = shared prognostic factor assumption; STC = simulated treatment
comparison \par
}

\vspace{3mm}
\noindent \textbf{Fig. C.2} Bias for Log Odds Ratio (n=300)

\includegraphics[scale=0.5]{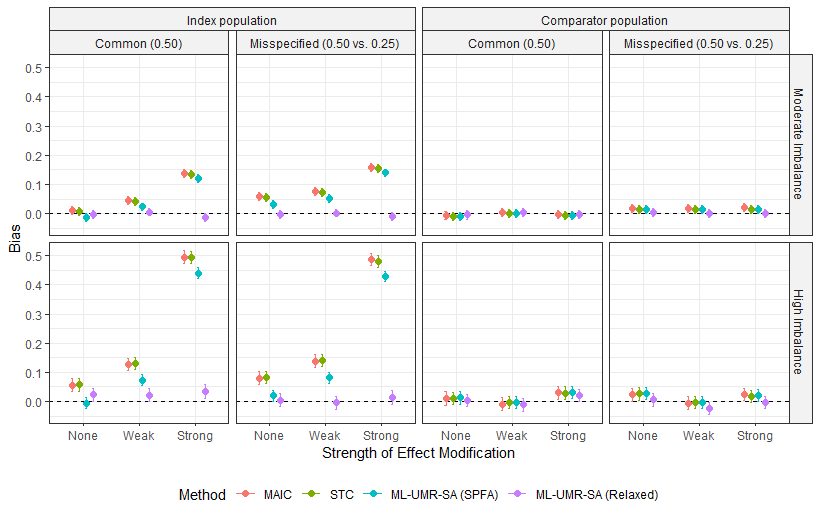}

{\footnotesize
\setlength{\baselineskip}{0.9\baselineskip}
\noindent Abbreviations: EM = effect modification; MAIC = matching-adjusted
indirect comparison; ML-UMR-SA = multilevel unanchored meta-regression;
SPFA = shared prognostic factor assumption; STC = simulated treatment
comparison \par
}

\vspace{3mm}
\noindent \textbf{Fig. C.3} Coverage of 95\% Credible Intervals for Log Risk Ratio
(n=300)

\includegraphics[scale=0.5]{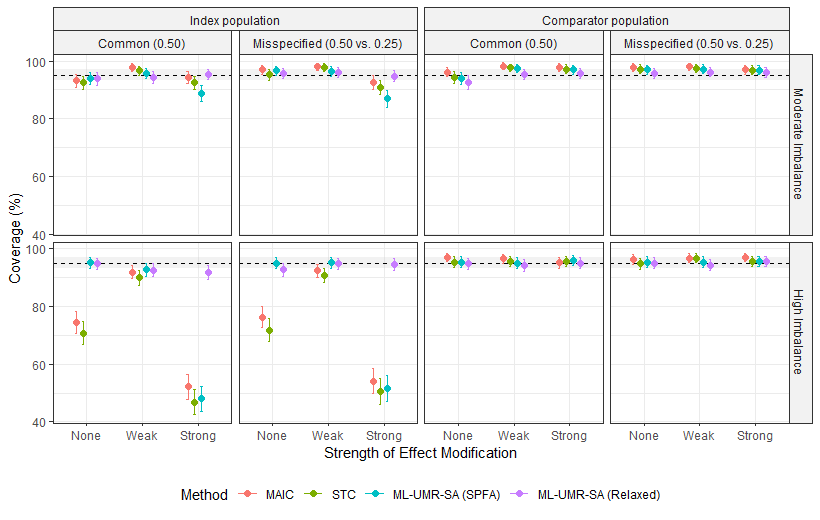}

{\footnotesize
\setlength{\baselineskip}{0.9\baselineskip}
\noindent Abbreviations: EM = effect modification; MAIC = matching-adjusted indirect comparison; ML-UMR-SA = multilevel unanchored meta-regression; SPFA = shared prognostic factor assumption; STC = simulated treatment comparison \par
}

\vspace{3mm}
\noindent \textbf{Fig. C.4} Coverage of 95\% Credible Intervals for Log Odds Ratio
(n=300)

\includegraphics[scale=0.5]{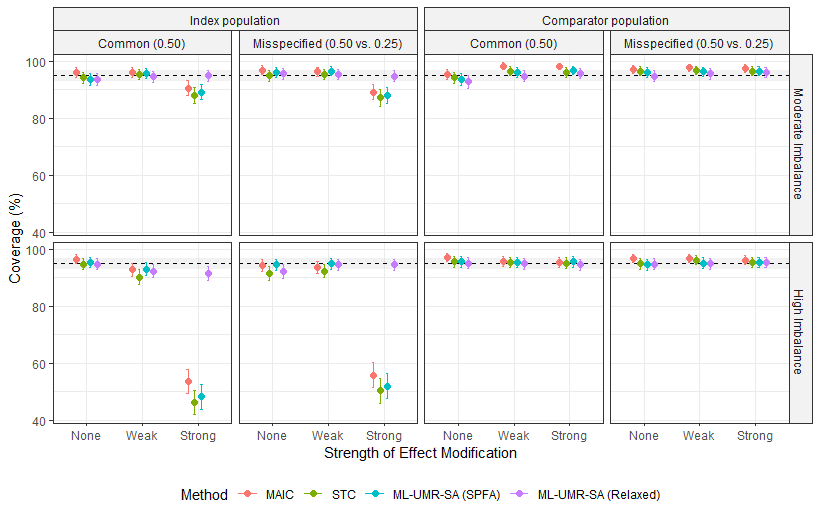}

{\footnotesize
\setlength{\baselineskip}{0.9\baselineskip}
\noindent Abbreviations: EM = effect modification; MAIC = matching-adjusted indirect comparison; ML-UMR-SA = multilevel unanchored meta-regression; SPFA = shared prognostic factor assumption; STC = simulated treatment comparison \par
}

\vspace{3mm}
\noindent \textbf{Fig. C.5} Effect of Sample Size on Bias in Index Population
(Common Correlation of 0.5)

\includegraphics[scale=0.5]{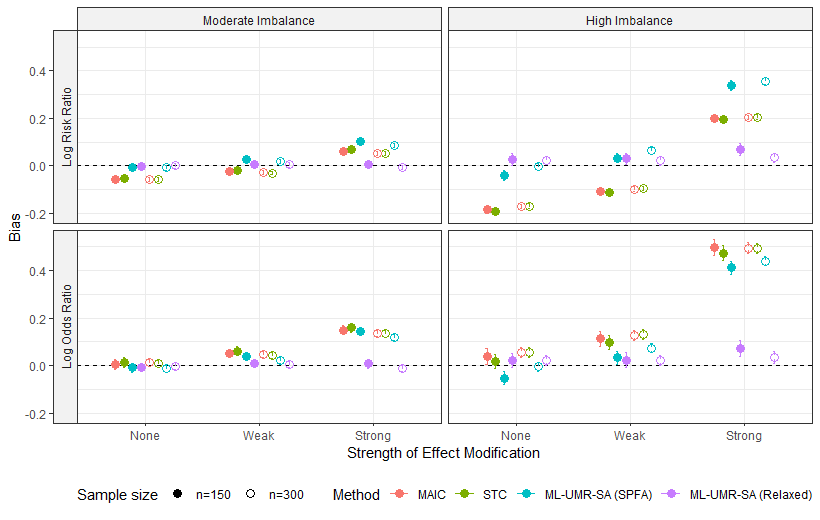}

{\footnotesize
\setlength{\baselineskip}{0.9\baselineskip}
\noindent Abbreviations: EM = effect modification; MAIC = matching-adjusted
indirect comparison; ML-UMR-SA = multilevel unanchored meta-regression;
SPFA = shared prognostic factor assumption; STC = simulated treatment
comparison \par
}

\hypertarget{c.8-additional-simulation-results}{%
\subsection{Sensitivity Analysis: Violation of Conditional Constancy Assumption}\label{c.8-additional-simulation-results}}

The primary simulation assumes that all prognostic factors responsible
for differences in absolute outcomes between study populations are
observed and appropriately adjusted for. To examine violation of this
conditional constancy assumption, we repeated all \(n = 150\) scenarios
while omitting \(X_{1}\) from the adjustment set for ML-UMR-SA, STC, and
MAIC. \(X_{1}\) remained part of the data-generating mechanism and
differed in prevalence between the index and comparator populations,
thereby acting as an unobserved prognostic factor in the fitted
analyses. All other aspects of the simulation were unchanged.

\vspace{3mm}
\noindent \textbf{Fig. C.6} Bias for Log Risk Ratio Excluding \(X_{1}\) from PAIC
Adjustment

\includegraphics[scale=0.5]{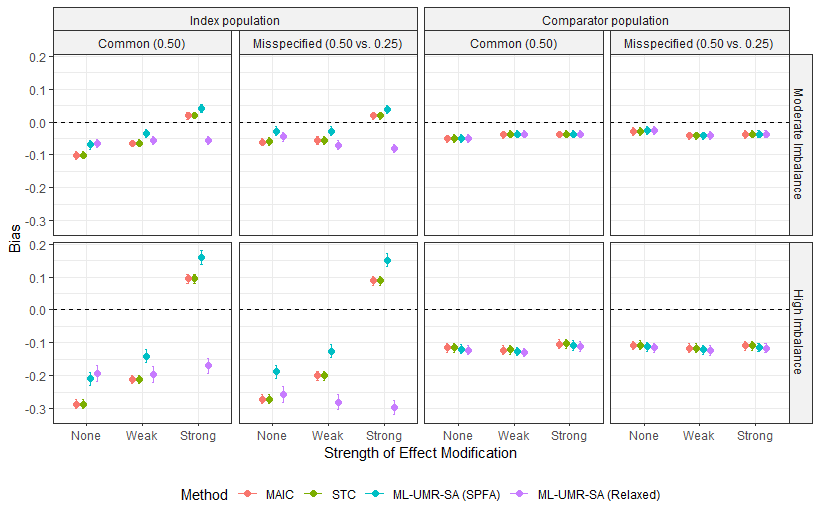}

\vspace{3mm}
\noindent \textbf{Fig. C.7} Bias for Log Odds Ratio Excluding \(X_{1}\) from PAIC
Adjustment

\includegraphics[scale=0.5]{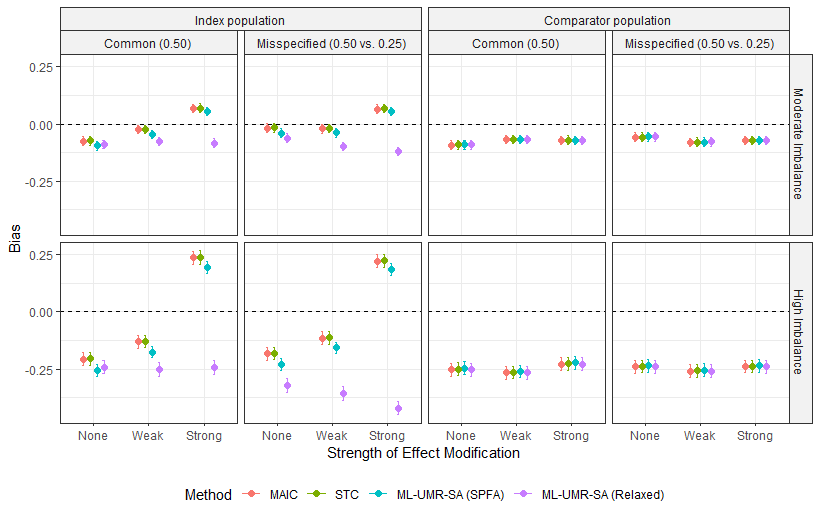}

\vspace{3mm}
\noindent \textbf{Fig. C.8} Coverage of 95\% Credible Intervals for Log Risk Ratio
Excluding \(X_{1}\) from PAIC

\includegraphics[scale=0.5]{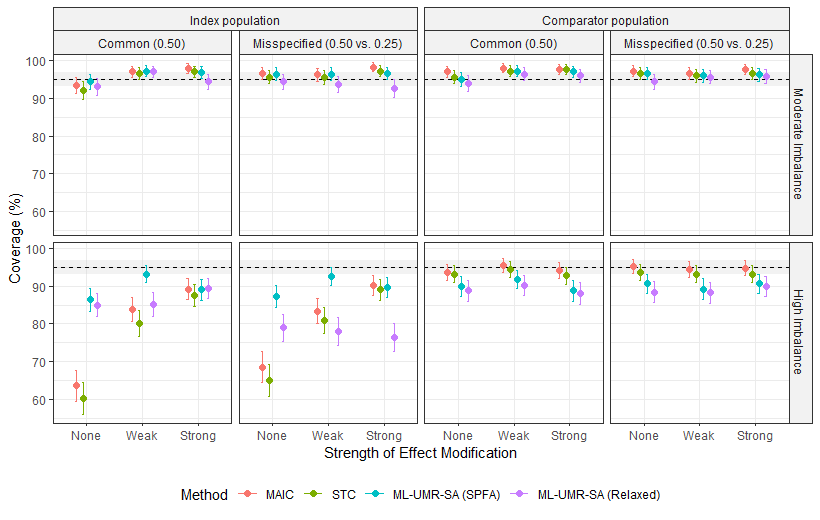}

\vspace{3mm}
\noindent \textbf{Fig. C.9} Coverage of 95\% Credible Intervals for Log Odds Ratio
Excluding \(X_{1}\) from PAIC

\includegraphics[scale=0.5]{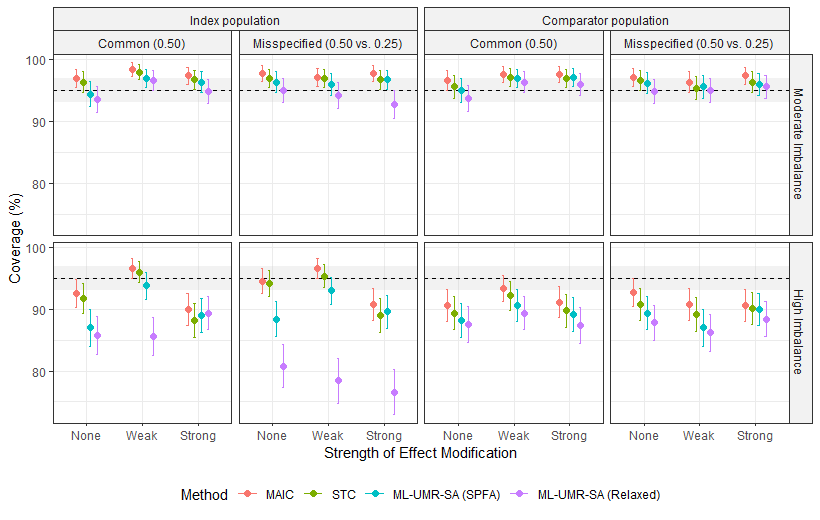}

\hypertarget{references-for-appendices-a-c}{%
\section{References for Appendices
A-C}\label{references-for-appendices-a-c}}

\hspace*{4mm} 1. Phillippo DM, Ades AE, Dias S, Palmer S, Abrams KR, Welton NJ. NICE
DSU Technical Support Document 18: Methods for population-adjusted
indirect comparisons in submissions to NICE. 2016.

2. Rosenbaum PR, Rubin DB. The central role of the propensity score in
observational studies for causal effects. Biometrika. 1983;70(1):41--55.

3. Hernán MA, Robins JM. Causal Inference: What If. Boca Raton: Chapman
\& Hall/CRC; 2020.

4. Phillippo DM, Ades AE, Dias S, Palmer S, Abrams KR, Welton NJ.
Methods for population-adjusted indirect comparisons in health
technology appraisal. Medical Decision Making. 2018;38(2):200-11.

5. Remiro-Azocar A, Heath A, Baio G. Parametric G-computation for
compatible indirect treatment comparisons with limited individual
patient data. Res Synth Methods. 2022 Nov;13(6):716-44.

6. Remiro-Azocar A, Heath A, Baio G. Model-based standardization using
multiple imputation. BMC Med Res Methodol. 2024 Feb 10;24(1):32.

7. Ren S, Ren S, Welton NJ, Strong M. Advancing unanchored simulated
treatment comparisons: A novel implementation and simulation study. Res
Synth Methods. 2024 Jul;15(4):657-70.

8. Snowden JM, Rose S, Mortimer KM. Implementation of G-computation on a
simulated data set: demonstration of a causal inference technique. Am J
Epidemiol. 2011 Apr 1;173(7):731-8.

9. Chandler C, Ishak KJ. Anchors Away: Navigating Unanchored Indirect
Comparisons With Multilevel Unanchored Meta-Regression. Value in Health.
2025;28(12):S498.

10. Chandler C, Ishak KJ. Reframing population-adjusted indirect
comparisons as a transportability problem: An estimand-based perspective
and implications for health technology assessment. arXiv preprint
arXiv:260217041. 2026.

11. Morris TP, White IR, Crowther MJ. Using simulation studies to
evaluate statistical methods. Stat Med. 2019 May 20;38(11):2074-102.

12. Phillippo DM, Dias S, Ades AE, Welton NJ. Multilevel network
meta-regression for population-adjusted treatment comparisons. Journal
of the Royal Statistical Society: Series A. 2020;183(3):1189-210.

13. Caro JJ, Ishak KJ. No head-to-head trial? simulate the missing arms.
Pharmacoeconomics. 2010;28(10):957-67.

14. Ishak KJ, Proskorovsky I, Benedict A. Simulation and matching-based
approaches for indirect comparison of treatments. Pharmacoeconomics.
2015 Jun;33(6):537-49.

15. Zhang L, Bujkiewicz S, Jackson D. Four alternative methodologies for
simulated treatment comparison: How could the use of simulation be
re-invigorated? Res Synth Methods. 2024 Mar;15(2):227-41.

16. Ishak KJ, Chandler C, Liu FF, Klijn S. A Framework for Reliable,
Transparent, and Reproducible Population-Adjusted Indirect Comparisons.
Pharmacoeconomics. 2025 May 5.

17. Signorovitch JE, Sikirica V, Erder MH, Xie J, Lu M, Hodgkins PS, et
al. Matching-adjusted indirect comparisons: a new tool for timely
comparative effectiveness research. Value Health. 2012
Sep-Oct;15(6):940-7.

18. Signorovitch JE, Wu EQ, Yu AP, Gerrits CM, Kantor E, Bao Y, et al.
Comparative effectiveness without head-to-head trials: a method for
matching-adjusted indirect comparisons applied to psoriasis treatment
with adalimumab or etanercept. Pharmacoeconomics. 2010;28(10):935-45.

\end{document}